\documentclass[secnumarabic, twocolumn, graphics,floatfix, superscriptaddress,tightenlines, aps, prb]{revtex4-2}
\usepackage[dvips]{graphicx}
\usepackage{amssymb,amsmath}
\usepackage{siunitx}
\usepackage{bm}
\usepackage{color}
\usepackage{multirow}
\usepackage{capt-of}

\begin{document}

\title{Dynamic polarization of nuclear spins by optically-oriented electrons and holes in lead halide perovskite semiconductors}

%\title{Dynamic spin polarization of nuclei via optically pumped localized electrons and holes in FA\textsubscript{0.9}Cs\textsubscript{0.1}PbI\textsubscript{2.8}Br\textsubscript{0.2} lead halide perovskite crystal}

\author{Mladen Kotur}
\affiliation{Experimentelle Physik 2, Technische Universit\"{a}t Dortmund, 44227 Dortmund, Germany}
\author{Pavel S. Bazhin}
\affiliation{Spin Optics Laboratory, St. Petersburg State University, 198504 St. Petersburg, Russia}
\author{Kirill V. Kavokin}
\affiliation{Spin Optics Laboratory, St. Petersburg State University, 198504 St. Petersburg, Russia}
\author{Nataliia E. Kopteva}
\affiliation{Experimentelle Physik 2, Technische Universit\"{a}t Dortmund, 44227 Dortmund, Germany}
\author{Dmitri R. Yakovlev}
\affiliation{Experimentelle Physik 2, Technische Universit\"{a}t Dortmund, 44227 Dortmund, Germany}
\affiliation{Ioffe Institute, Russian Academy of Sciences, 194021 St. Petersburg, Russia}
\author{Dennis Kudlacik}
\affiliation{Experimentelle Physik 2, Technische Universit\"{a}t Dortmund, 44227 Dortmund, Germany}
\author{Manfred Bayer}
\affiliation{Experimentelle Physik 2, Technische Universit\"{a}t Dortmund, 44227 Dortmund, Germany}
\affiliation{Research Center FEMS, Technische Universit\"at Dortmund, 44227 Dortmund, Germany}

\begin{abstract}
A theory of dynamic polarization of the nuclear spin system via optically-oriented charge carriers in lead halide perovskites is developed and compared with the experiments performed on a FA$_{0.9}$Cs$_{0.1}$PbI$_{2.8}$Br$_{0.2}$ crystal. The spin Hamiltonians of the electron and hole hyperfine interaction  with the nuclear spins of lead and halogen are derived. The hyperfine interaction of the halogen spins with charge carriers is shown to be anisotropic and depending on the position of the halogen nucleus in the cubic elementary cell. The quadrupole splitting is absent for the lead spins, but plays an important role for the halogen spins and affects their dynamic polarization by charge carriers. 
%Expressions for the Overhauser fields, created by dynamically polarized nuclear spins and acting on the spins of electrons and holes, are obtained.
The Overhauser fields of the dynamically polarized nuclei are calculated as functions of the tilting angle of an external magnetic field and compared with the experimentally measured angular dependence of the Hanle effect. The comparison of the theoretical model with the experimental data reveals an enhanced spin polarization of the lead nuclei, whose mean spin exceeds several times the mean spins of localized electrons and holes. This unexpectedly strong spin polarization is explained by the interaction of the lead nuclei with excitons having a high degree of spin orientation due to their short lifetime after excitation by circularly-polarized light. The dynamic polarization of the quadrupole-split halogen spins manifests itself via the magnetic field they produce at  the lead nuclei. This field maintains the magnetization of the lead nuclei at zero external magnetic field. The dynamics of the nuclear spin polarization is measured under optical pumping and in the dark, yielding a nuclear spin-lattice relaxation time on the order of 10 seconds.
\end{abstract}

%\pacs{} 
\maketitle


\section{Introduction}
\label{sec:intro}

Among solids, semiconductors are unique in containing localized states of charge carriers that expand over a great number of elementary cells of their crystal lattice. The spins of the localized electrons and holes couple with thousands to millions of lattice nuclei via the hyperfine interaction. Once these nuclei acquire an average spin polarization, they act on the spins of the charge carriers as an effective magnetic field. Since localized carriers can efficiently exchange their spin with the nuclei, nuclear spin polarization in semiconductors can be easily created by pumping the mean spin of carriers with circularly polarized light. In this way, an intricate spin system can be created that demonstrates complex, but controllable dynamics, potentially usable for spintronics and quantum information applications. The  dynamic polarization of nuclear spins and their coupling to the radiative states of charge carriers serve as the basis for various optically-detected nuclear magnetic resonance (ODNMR) techniques~\cite{dyakonov2017}, including zero- and ultra-low field ODNMR~\cite{litvyak2021warm,litvyak2024nuclear}, giving access to subtle details of spin-spin interactions and material tensors, which ultimately reveal the atomic structure of the crystal.
%%%%%%%%%%%%%%%%%%%%%%%%%%%%%%%%%%%%%%%%%%%%%%%%%%

Lead halide perovskite semiconductors are emerging materials, which demonstrate remarkable optical and optoelectronic properties~\cite{Vinattieri2021_book,Vardeny2022_book}. Furthermore, they show attractive spin properties, which can be accessed by the optical and magneto-optical experimental techniques  well-established in the spin physics of semiconductors~\cite{dyakonov2017,meier2012optical}. Among them are polarized photoluminescence, spin-flip Raman scattering, and time-resolved Faraday/Kerr rotation~\cite{kopteva2023giant,belykh2019coherent,kirstein2022nc,kirstein2022lead}. These techniques allow one to address the coherent spin dynamics of electrons and holes, and to measure their spin coherence, spin dephasing, and longitudinal spin relaxation times, as well as Land\'e $g$-factors. In these experiments, also the effects provided by the hyperfine interaction of the charge carrier spins with the nuclear spin system were identified. The dynamic nuclear polarization (DNP) induced by the spin-oriented carriers was reported for several perovskite crystals: FA$_{0.9}$Cs$_{0.1}$PbI$_{2.8}$Br$_{0.2}$~\cite{kirstein2022lead}, MAPbI$_3$~\cite{kirstein2022mapi}, FAPbBr$_3$~\cite{Kirstein2023DNSS}, and CsPbBr$_3$~\cite{belykh2019coherent}. The nuclear spin fluctuations play an important role in the spin dynamics of localized electrons and holes at cryogenic temperatures, as we have shown for FA$_{0.9}$Cs$_{0.1}$PbI$_{2.8}$Br$_{0.2}$  crystals using the Hanle and polarization recovery effects~\cite{kudlacik2024optical}. In FAPbBr$_3$ crystals, the suppression of the nuclear spin fluctuations due to  creation of a squeezed dark nuclear spin state was demonstrated~\cite{Kirstein2023DNSS}. Recently, we have reported on the hole hyperfine interaction with the nuclear spin fluctuations in CsPbBr$_3$ and CsPb(Cl,Br)$_3$ nanocrystals~\cite{Meliakov2024} and on the electron hyperfine interaction in CsPbI$_3$ nanocrystals~\cite{Meliakov2025}. 

In contrast to conventional III-V and II-VI direct-gap semiconductors, lead halide perovskite semiconductors have a simple structure of the conduction and valence bands with spins of 1/2. In the perovskites the holes, as well as the electrons, remain spin-polarized for a relatively long time (in the nanosecond range) after excitation with circularly polarized light. In lead halide perovskites, the valence band is significantly contributed by the $s$-orbitals of the Pb-ions, while the conduction band is formed mostly by the $p$-orbitals of the Pb-ions with a small admixture of the $s$-orbitals of the halogens. As a result, the hole-nuclei hyperfine interaction is about 5 times stronger than that of the electron-nuclei~\cite{kirstein2022lead}, which is opposite to the behavior in conventional semiconductors.

%%%%%%%%%%%%%%%%%%%%%%%%%%%%%%%%%%%%%%%%%%%%%%

In this paper, we use optical excitation with circularly polarized light of constant helicity to pump nuclear spins in a single crystal of FA$_{0.9}$Cs$_{0.1}$PbI$_{2.8}$Br$_{0.2}$ lead halide perovskite. The resulting nuclear spin polarization is detected via its back action on the spin polarization of electrons and holes, as revealed by the magnetic field dependence of the photoluminescence (PL) polarization. This dependence has to be compared with a reference dependence obtained under conditions where the nuclear spins remain unpolarized, which is conveniently realized by using helicity-alternated optical excitation~\cite{dyakonov2017,meier2012optical}. In the absence of nuclear spin polarization, the PL polarization is determined by the interaction of the electron and hole spins with nuclear spin fluctuations. These fluctuations create effective magnetic fields, whose strength and direction vary from one carrier localization site to another. At zero or in weak external magnetic fields, the PL polarization is determined by the competing processes of carrier spin precession in random nuclear fields, carrier hopping to other localization sites, and recombination. As shown in Ref.~\onlinecite{kudlacik2024optical}, the Larmor precession periods of both the electron and hole spins in the FA$_{0.9}$Cs$_{0.1}$PbI$_{2.8}$Br$_{0.2}$ crystal are shorter than the hopping and recombination times. Under these conditions, the external magnetic field starts to influence the PL polarization once it becomes stronger than the root mean squared (rms) fluctuation field. If the external field is applied along the $k$-vector of the excitation light, the carrier spin becomes less affected by the random nuclear fields, and the PL polarization is enhanced (the polarization recovery effect). If the external field is applied in the perpendicular direction, it adds up to the depolarizing effect of the nuclear spin fluctuations, and the PL polarization decreases (the Hanle effect). 

%The polarization recovery and Hanle curves have  similar widths close to the rms fields of nuclear fluctuations. Since electrons and holes have different $g$-factors and localization radii, the rms nuclear fields for the two types of carriers are different. As a result, polarization recovery and Hanle curves are composed of contours of different widths. Two of them are associated with electrons and holes. There is also a third, narrow contour, tentatively attributed to weakly localized holes []. 

In order to polarize nuclear spins by optical pumping, the helicity of the excitation light should be constant, and an external magnetic field should be applied to stabilize the accumulated nuclear spin polarization. The field should have a nonzero projection on the k-vector of light and should be stronger than the local fields created by the nuclear spins at each other location. In this work, we apply the magnetic field at an angle to the $k$-vector of circularly-polarized laser beam. In this geometry, the nuclear spin polarization builds up parallel or antiparallel to the magnetic field, depending on the excitation helicity ($\sigma^+$ or $\sigma^-$). Correspondingly, spin-polarized nuclei create an effective magnetic field (the Overhauser field) along the external field or opposite to it, thereby shifting the observed Hanle curve by the value of the Overhauser field. Similarly, in Faraday geometry the Overhauser field adds to the external field or subtracts from it, thereby modifying the polarization recovery effect. %\cD{Kirill, above we comment only tilted field case and appearance in Hanle, but in fact we demonstrate that also for Faraday geometry and PR curve as well.}

The paper is organized as follows. We start in Sec.~\ref{sec:theory} with the theoretical consideration of the nuclear spin system in lead halide perovskite semiconductors, considering the specifics of the lead and halogen nuclei and their interaction with each other. Then we turn to the hyperfine interaction of the nuclear spins with electrons and holes and consider the dynamic nuclear polarization in the regimes of strong and weak magnetic fields. In Section~\ref{sec:experimentals}, technological details of the studied FA$_{0.9}$Cs$_{0.1}$PbI$_{2.8}$Br$_{0.2}$ single crystal belonging to the lead halide perovskite semiconductors are given. Also the experimental techniques of optical spin orientation in different geometries of the external magnetic field are described. The experimental results on the optical orientation of electrons and holes and the manifestation of the hyperfine interaction in the Hanle and polarization recovery effects are presented in Sec.~\ref{sec:exp}. Here, dynamic nuclear polarization is achieved and exploited for measuring the spin-relaxation time of the nuclear spin system. In Sec.~\ref{sec:discussion} we discuss and analyze the experimental results in the frame of the developed theory and available literature data.

\section{Theory}
\label{sec:theory}

\subsection{Introduction to the nuclear spin dynamics in perovskites}
\label{sec:theory_intro}

The non-equilibrium spin polarization of charge carriers, created by the optical spin orientation in semiconductors, can be transferred to the spins of the lattice nuclei by the hyperfine interaction \cite{OOChapter5}. As a result of this process, known as dynamic nuclear polarization (DNP), the nuclear spin system (NSS) acquires a certain magnetization. If an external magnetic field is applied, the projection of the nuclear magnetization onto this field can persist during the nuclear spin-lattice relaxation time $T_1$, which is orders of magnitude longer than the spin lifetimes of charge carriers. The polarized nuclear spins create effective magnetic fields (Overhauser fields), acting on the spins of electrons and holes via the hyperfine interaction. In this way, the DNP affects the spin dynamics of the carriers, which can be observed in time-resolved (spin beats under pulsed optical excitation) as well as time-integrated (Hanle and polarization recovery effects) experiments \cite{SPSDyakonov}. The form and the magnitude of the DNP effects depend on the strength and the symmetry of the spin interactions in the semiconductor. These interactions include: (i) the quadrupole interaction with electric field gradients at the nuclei positions (for the nuclei with the spin value greater than $1/2$), (ii) the dipole-dipole interaction between the nuclear magnetic moments, (iii) the above-mentioned hyperfine interaction with the carrier spins, (iv) the spin-orbit and exchange interactions of the charge carriers, and (v) the indirect coupling of the nuclear spins via electrons.

The lead halide perovskites differ from conventional III-V and II-VI semiconductors in several important aspects concerning the electron-nuclear spin dynamics. First of all, a difference is the band structure with a simple valence band and the hole spin $1/2$. As a result, both types of charge carriers demonstrate efficient optical orientation of the spins and contribute to the DNP \cite{belykh2019coherent}. Also, due to the specific crystal structure, the spins of the halogen nuclei experience a strong quadrupole splitting even in the cubic phases of perovskite crystals \cite{piveteau2020bulk}. In this section, we consider the relevant spin interactions and present a theory of dynamic nuclear polarization for the cubic phase of lead halide perovskite semiconductors.

In the lead halide perovskite semiconductors, the nuclei with non-zero spin and magnetic moment, which experience hyperfine coupling with the charge carriers, are given by $^{207}\text{Pb}$ having the spin $I = 1/2$ and natural abundance of 22\%, and the halogens ($^{127}\text{I}$ with $I = 5/2$, $^{35/37}\text{Cl}$ and $^{79/81}\text{Br}$ with $I=3/2$) all having 100\% natural abundance of the magnetic isotopes. There are also $^{133}\text{Cs}$ with $I=5/2$ as well as $^{14}\text{N}$ and $^{1}\text{H}$ in the organic cations, that apparently do not interact with carriers, but are coupled with the lead and halogen nuclei by the magnetic dipole-dipole interaction.

\subsubsection{Quadrupole interaction of halogen spins}
\label{sec:theory_halogens}

In the perovskite lattice, the halogen nuclei are subject to a large uniaxial electric field gradient (EFG) with the main axis along the Pb-hal-Pb bond line, where "hal" is the halogen. Since all halogen nuclei have spins $I>1/2$ and non-zero quadrupole moments, their spin multiplets experience a quadrupole splitting into a series of Kramers doublets with the spin projections $M_Q$ onto the Pb-hal-Pb axis equal to  $\pm 1/2$, $\pm 3/2$, or $\pm 5/2$. According to nuclear quadrupole resonance data \cite{piveteau2020bulk}, the quadrupole splitting of $^{79/81}\text{Br}$ in bulk orthorhombic $\text{CsPbBr}_3$ is 67 MHz, while for $^{127}\text{I}$ in bulk orthorhombic $\text{CsPbI}_3$ it is approximately 80 MHz between the $M_Q=\pm 1/2$ and $M_Q=\pm 3/2$ doublets and 160 MHz between the $M_Q=\pm 3/2$ and $M_Q=\pm 5/2$  doublets.

These values correspond to Zeeman splittings of the nuclear spin levels in a magnetic field of more than 10 Tesla strength. Therefore, the weak magnetic fields (below 100 mT) used in our experiments do not couple the spin states belonging to different Kramers doublets. The splitting of the two states within each doublet depends on the magnetic field orientation with respect to the EFG axis. Each of the doublets can, therefore, be treated as a spin $1/2$ with an anisotropic $g$-factor \cite{OOChapter5}, so that their Zeeman interaction can be written as
\begin{equation}
    \hat{H}_{Z,IM}=\frac{1}{2}\hbar\gamma_N\left( g_1^{IM_Q}B_1\hat\sigma_x+g_2^{IM_Q}B_2\hat\sigma_y+g_3^{IM_Q}B_3\hat\sigma_z\right)\,.
    \label{eq:eq1}
\end{equation}
Here, $\hbar$ is the reduced Plank constant, $\gamma_N$ is the gyromagnetic ratio of the nucleus $N$, $g_i^{IM_Q} $ are the principal values of the anisotropic $g$-factor, defined for the $M_Q$ doublet of a nucleus with the spin $I$, $\hat\sigma_{x,y,z}$ are the Pauli matrices defined on the pair of states $|I,\pm M_Q\rangle$, $B_3$ is the magnetic field component along the main EFG axis, while  $B_1$ and $B_2$ are the field components along the two perpendicular principal axes of the EFG tensor.

The quadrupole Hamiltonian reads
\begin{equation}
    {{\hat{H}}_{Q}}={{A}_{Q}}\left( 3\hat{I}_{z}^{2}-I\left( I+1 \right)+\frac{\eta }{2}\left( \hat{I}_{+}^{2}+\hat{I}_{-}^{2} \right) \right),
    \label{eq:eq2}
\end{equation}
where $\hat{I}_{\pm} = \hat{I}_x\pm i\hat{I}_y$, $\hat{\vec{I}} \!=\!\left(\hat{I}_x, \hat{I}_y, \hat{I}_z\right)^T$ is the vector spin operator, $A_Q$ is a characteristic energy and $\eta$ is the non-axiality parameter of the EFG tensor \cite{abragam1961principles}. Treating the non-axiality as a perturbation, one obtains the expressions for the effective $g$-factor components, listed in Table \ref{tab:gfactors}.
\begin{table}[b]
\caption{\label{tab:gfactors} Effective $g$-factor components for nuclei with spins $I=3/2$ and $I=5/2$.}
\footnotesize
\begin{ruledtabular}
\begin{tabular}{>{\centering\arraybackslash}p{0.12\linewidth}>{\centering\arraybackslash}p{0.19\linewidth}>{\centering\arraybackslash}p{0.29\linewidth}>{\centering\arraybackslash}p{0.36\linewidth}}
    &$M_Q=\pm 1/2$& $M_Q=\pm 3/2$& $M_Q=\pm 5/2$\\ \hline
    \multirow{2}{*}{$I=3/2$} & $g_3=1$ & $g_3 = 3$ & \multirow{2}{*}{---} \\ 
    & $g_1=g_2=2$&$g_1=g_2=g_\perp=\eta$&\\ \hline
    \multirow{2}{*}{$I=5/2$} & $g_3=1$ & $g_3 = 3$ & $g_3 = 5$ \\ 
     &$g_1=g_2=3$&$g_1=g_2=g_\perp=4\eta$& $g_1=g_2=g_\perp=\frac{10}{27}\eta^2$\\
\end{tabular}
\end{ruledtabular}
\end{table}

In the tetragonal phase of $\text{MAPbI}_3$, $\eta\approx0.01$ \cite{xu1991molecular} and one could expect it to be even smaller in the cubic lead halide perovskites. Therefore, the transverse components of the effective $g$-factor are less than 0.04 in the absolute value for all doublets except the $M_Q=\pm 1/2$ one. Under these conditions, the $M_Q=\pm 3/2$ and $M_Q=\pm 5/2$ doublets should demonstrate the anomalously slow spin dynamics \cite{OOChapter5, artemova1985polarization}. Indeed, the correlation function of the nuclear magnetic moment component along the Pb-hal-Pb axis decays on the time scale of $T_Q \sim T_2/g^2_\perp$. Here $T_2\sim 10^{-4}$ s is the spin-spin relaxation time of the nuclei in the absence of a quadrupole splitting and $g_\perp$ is the principal value of the anisotropic $g$-factor perpendicular to the EFG main axis.  With $g_\perp < 0.04$ , one would expect $T_Q$ to last at least hundreds of milliseconds, i.e. to fall into the range of the spin-lattice relaxation times $T_1$.

\subsubsection{\label{sec:dipole-dipoleInteraction}Magnetic dipole-dipole interaction of nuclear spins}
\label{sec:theory_dip_dip}

The dipole-dipole interaction of the nuclear magnetic moments governs the nuclear spin dynamics in low external magnetic fields. Its effect is conventionally modeled by introducing a local nuclear magnetic field. The dipole-dipole interaction of two nuclear spins with gyromagnetic ratios $\gamma_1$, $\gamma_2$ and spins $I_1$, $I_2$ can be described by:
\begin{equation}
    \widehat{H}_{ss}=\frac{\hbar^2\gamma _1\gamma_2}{r_{12}^3}\left( \left(\hat{\vec{I}}_1\hat{\vec{I}}_2 \right)-\frac{3\left( {{{\hat{\vec{I}}}}_{1}}{{\vec{r}}_{12}} \right)\left( {{{\hat{\vec{I}}}}_{2}}{{\vec{r}}_{12}} \right)}{r_{12}^{2}} \right),
    \label{eq:eq3}
\end{equation}
where $\vec{r}_{12}$ is the radius-vector between  two nuclei. Then for the local magnetic field exerted by the second spin on the first one can write:
\begin{equation}
{{\widehat{\vec{B}}}_{12}}=\frac{\hbar {{\gamma }_{2}}}{r_{12}^{3}}\left( \frac{3\left( {{{\hat{\vec{I}}}}_{2}}{{\vec{r}}_{12}} \right)}{r_{12}^{2}}{{\vec{r}}_{12}}-{{{\hat{\vec{I}}}}_{2}} \right).
    \label{eq:eq4}
\end{equation}
The total dipolar field at a given nuclear site $\widehat{B}_1$ is calculated by summing Eq.~\eqref{eq:eq4} over all lattice nuclei. If the nuclei are not spin-polarized, quantum mechanical averaging over the isotropic spin distribution results in a vanishing first moment of the local field. Consequently, the observable effects arise from the second moment of the field $\langle B^2_1 \rangle$, which is determined by
\begin{equation}
    \left\langle B_{1}^{2} \right\rangle =\operatorname{Tr}\!\left(\rho{\widehat{B}_1^{2}} \right)\!=\!\sum\limits_{s}{2{{I}_{s}}\left( {{I}_{s}}+1 \right){{\left( \hbar {{\gamma }_{s}} \right)}^{2}}\sum\limits_{i}{\frac{1}{r_{i,s}^{6}}}}\,.
    \label{eq:eq5}
\end{equation}
%
%where index $s$ runs over all magnetic isotopes in $\text{FAPbI}_3$ and $i$ runs over all nuclei of the isotope $s$. To estimate the contribution from formamidinium one can assume that the spins of 5 hydrogen nuclei ($I = 1/2$, $\gamma_H~=~267600$  rad $\text{mT}^{-1}\text{s}^{-1}$) and 2 nitrogen nuclei ($I = 1$, $\gamma_N~=~19320$ rad $\text{mT}^{-1}\text{s}^{-1}$) at the location of the organic complex are not correlated. The Eq. \eqref{eq:eq5}, with the sum taken over I, FA and Pb nuclei, then gives $\langle B^2_{\text{Pb}}\rangle \approx 0.051$~$\text{mT}^2$, and  $\langle B^2_{\text{I}}\rangle \approx 0.037$ $\text{mT}^2$ for the mean squared local field at the position of the Pb and I nuclei. These values correspond to the local fields $\sqrt{\langle B^2_{\text{Pb}}\rangle} \approx 0.226$ mT and $\sqrt{\langle B^2_{\text{I}}\rangle} \approx 0.193$ mT. 
%
Here the index $s$ runs over all magnetic isotopes and $i$ runs over all nuclei of the isotope $s$. Using this equation, one can estimate the local field $B_L=\sqrt{\langle B^2_1\rangle}$ at the given nucleus. The calculated local field at the Pb and halogen nuclei for different perovskites is listed in Table \ref{tab:DipolarFields}. The contribution of the Pb, halogen and organic complexes or the Cs nuclei was taken into account. To estimate the contribution of the organic complexes, it was assumed that the spins of the hydrogen and nitrogen nuclei are located in the middle of the cubic cells and are not correlated. 

\begin{table*}[t]
\caption{\label{tab:DipolarFields} Local fields acting on the Pb and halogen nuclei and constants $L_1$, $L_2$, $L_3$ included in Eqs. \eqref{eq:eq6}, \eqref{eq:eq7}, \eqref{eq:eq8} for different perovskites.}
\begin{ruledtabular}
\begin{tabular}{>{\centering\arraybackslash}m{0.22\linewidth}>{\centering\arraybackslash}p{0.08\linewidth}>{\centering\arraybackslash}p{0.08\linewidth}>{\centering\arraybackslash}p{0.08\linewidth}>{\centering\arraybackslash}p{0.08\linewidth}>{\centering\arraybackslash}p{0.08\linewidth}>{\centering\arraybackslash}p{0.08\linewidth}>{\centering\arraybackslash}p{0.08\linewidth}>{\centering\arraybackslash}p{0.08\linewidth}}
    & $\text{FAPbI}_3$ &$\text{FAPbBr}_3$ &$\text{MAPbI}_3$ &$\text{MAPbBr}_3$ &$\text{MAPbCl}_3$ &$\text{CsPbI}_3$ &$\text{CsPbBr}_3$ &$\text{CsPbCl}_3$ \\
    \hline
    {Local field at Pb (mT)} & 0.23\rule{0pt}{9pt} & 0.25 & 0.25 & 0.27 & 0.23 & 0.20 & 0.20 & 0.10 \\
    \hline
    {Local field at halogen (mT)} & 0.20\rule{0pt}{9pt} & 0.23 & 0.22 & 0.26 & 0.28 & 0.11 & 0.11 & 0.08 \\
    \hline
    $L_1$ (mT)& 0.066\rule{0pt}{9pt} & 0.103 & 0.069 & 0.106 & 0.044 & 0.069 & 0.109 & 0.045\\
    \hline
    $L_2$ (mT)& 0.019\rule{0pt}{9pt} & 0.030 & 0.020 & 0.030 & 0.013 & 0.020 & 0.031 & 0.013\\
    \hline
    $L_3$ (mT)& 0.009\rule{0pt}{9pt} & 0.015 & 0.010 & 0.015 & 0.006 & 0.010 & 0.016 & 0.007\\
\end{tabular}
\end{ruledtabular}
\end{table*}

The presence of spin-polarized nuclei can lead to the emergence of a net magnetic field aligned along the polarization axis, although in the case of cubic symmetry, this field still averages to zero. We estimated the fields which act on the Pb and halogen nuclei. The main contribution to the field is given by the halogen nuclei. Suppose that a non-zero polarization of the halogen nuclei has been created along the cubic axis [001]. The presence of strong quadrupole interactions results in unequal polarization of the nuclei, located in the hal-Pb-hal chains aligned with different crystallographic directions. We will mark their contributions by a superscript at the average spin, indicating the axis along which the hal-Pb-hal chain is aligned, so that $\langle I^Z\rangle\ne \langle I^Y\rangle=\langle I^X\rangle=\langle I^{X/Y}\rangle$. This difference is responsible for the non-zero magnetic field experienced by the lead and halogen nuclei. 

Then the field acting on the Pb nuclei is given by:
\begin{equation}
\begin{split}
       &\left\langle B_{Pb} \right\rangle \!=\!\operatorname{Tr}\left( \rho \widehat{B}_{Pb} \right)\!=\!\frac{\hbar \gamma _h}{a^3}\Bigg[ \left\langle {I^Z} \right\rangle \sum\limits_i{\frac{3\frac{(\vec{r}_{Pb,i})^2_z}{r_{Pb,i}^2}-1}{\left(r_{Pb,i}/a\right)^3}} +\\
   &+\!\left\langle I^{X/Y} \right\rangle \!\!\sum\limits_i{\frac{3\frac{(\vec{r}_{Pb,i})^2_z}{r_{Pb,i}^2}-1}{\left(r_{Pb,i}/a\right)^3}}\Bigg]\!\!=\!L_1\!\times \!\left[ \left\langle I_z^Z\right\rangle\! -\!\left\langle I_{z}^{X/Y} \right\rangle  \right],
\end{split}
\label{eq:eq6}
\end{equation}
%
%\cN{what do $Z_{Pb}$, $a$, $r_{Pb}$, $i$ mean?} 
where the first sum is taken over the halogen nuclei located in the hal-Pb-hal chains aligned along the $Z$-axis, while the second sum includes the remaining halogen nuclei, $a$ is the lattice constant, $\vec{r}_{\text{Pb},i}$ is the radius vector between a Pb nucleus and the $i$-th halogen nucleus. $L_1$ (as well as $L_2$, $L_3$ in Eqs. \eqref{eq:eq7}, \eqref{eq:eq8}) is the constant pertinent to the specific perovskite. They are listed in Table \ref{tab:DipolarFields}.

For the field acting on the halogen nuclei, there are two cases: (i) halogen nuclei located in the hal-Pb-hal chains aligned along the $Z$-axis, and (ii) halogen nuclei located in the hal-Pb-hal chains aligned along the $X$ or $Y$ axis. In the first case:
\begin{equation}
\begin{split}
    &\left\langle {{B}_{h,Z}} \right\rangle =\frac{\hbar {{\gamma }_{h}}}{{{a}^{3}}}\left[ \left\langle I_{{}}^{X/Y} \right\rangle \sum\limits_{i}{\frac{3\frac{(\vec{r}_{h,i})^2_z}{r_{h,i}^{2}}-1}{{{\left( {{{r}_{h,i}}}/{a} \right)}^{3}}}} \right]=\\&L_2\times \left\langle I_{{}}^{X/Y} \right\rangle,
\end{split}
    \label{eq:eq7}
\end{equation}
where the sum is taken over the halogen nuclei located in the hal-Pb-hal chains aligned along the $X$ and $Y$ axes, $\vec{r}_{\text{h},i}$ is the radius vector between a halogen nucleus in the hal-Pb-hal chain aligned along the $Z$ axis and the $i$-th nucleus in the sum. In the second case:
\begin{equation}
\begin{split}
  &\left\langle B_{h,X/Y} \right\rangle =\frac{\hbar\gamma_h}{a^3}\Bigg[\left\langle I^Z \right\rangle \sum\limits_i{\frac{3\frac{(\vec{r}_{h,i})^2_z}{r_{h,i}^2}-1}{\left( r_{h,i}/a \right)^3}}+\\&+\!\left\langle I^{X/Y}\! \right\rangle \!\!\sum\limits_i{\frac{3\frac{(\vec{r}_{h,i})^2_z}{r_{h,i}^2}-1}{\left( r_{h,i}/a \right)^3}} \Bigg]\!\!=\!L_3\!\times\!\! \left[ \left\langle I_{z}^{Z} \right\rangle\!-\!2\left\langle I_{z}^{X/Y} \right\rangle  \right]\!,
\end{split}
\label{eq:eq8}
\end{equation}
where the sums run over the same set of nuclei as in Eq. \eqref{eq:eq6}.

\subsection{Theory of hyperfine interaction}
\label{sec:theory_hyperfine}

The Bloch amplitudes of electrons and holes at the $R$ point of the Brillouin zone in the lead halide perovskites are linear combinations of the several electron orbitals of the constituent atoms. Among them the $s$ and $p$ atomic orbitals of lead and the halogens are known for their considerable hyperfine coupling with the corresponding nuclear spins. In particular, in lead iodide perovskites the conduction band bottom (cbb) is strongly contributed by the Pb(6$p$) states with an admixture of the I(5$s$) orbitals, while the valence band top (vbt) states are linear combinations of the Pb(6$s$) and I(5$p$) orbitals with comparable weights \cite{boyer2016symmetry, targhi2018mapbi3, Meliakov2025}. %\cD{Kirill, shall we cite here our just submitted paper on NCs with Nastoklon calculations?} 
Taking into account the spin-orbit interaction, the cbb and vbt Bloch amplitudes are~\cite{kirstein2022lead}:
\begin{equation}
\begin{split}
   &u_{cbb}^{+1/2}=-\frac{{{C}_{cp}}}{\sqrt{3}}\left[ \left| {{Z}_{Pb}} \right\rangle \uparrow +\left( \left| {{X}_{Pb}} \right\rangle +i\left| {{Y}_{Pb}} \right\rangle  \right)\downarrow  \right]-\\&-\frac{{{C}_{cs}}}{\sqrt{3}}\left[ \left| {{S}_{hal,Z}} \right\rangle \uparrow +\left( \left| {{S}_{hal,X}} \right\rangle +i\left| {{S}_{hal,Y}} \right\rangle  \right)\downarrow  \right]+u_{c,res}^{+1/2} \,,\\
  &u_{cbb}^{-1/2}=\;\;\;\;\frac{{{C}_{cp}}}{\sqrt{3}}\left[ \left| {{Z}_{Pb}} \right\rangle \downarrow -\left( \left| {{X}_{Pb}} \right\rangle -i\left| {{Y}_{Pb}} \right\rangle  \right)\uparrow  \right]+\\&+\frac{{{C}_{cs}}}{\sqrt{3}}\left[ \left| {{S}_{hal,Z}} \right\rangle \downarrow -\left( \left| {{S}_{hal,X}} \right\rangle -i\left| {{S}_{hal,Y}} \right\rangle  \right)\uparrow  \right]+u_{c,res}^{-1/2} \,, \\ 
\end{split}
\label{eq:eq9}
\end{equation}
\begin{equation}
\begin{split}
  & u_{vbt}^{+1/2}\!\!=\!\!\left[ {{C}_{vs}}\!\left| {{S}_{Pb}} \right\rangle \!+\!\frac{{{C}_{vp}}}{\sqrt{3}}\!\left( \left| {{X}_{hal,X}} \right\rangle \!+\!\left| {{Y}_{hal,Y}} \right\rangle \!+\!\left| {{Z}_{hal,Z}} \right\rangle  \right) \right]\!\!\uparrow\!+\\& +u_{v,res}^{+1/2} \,, \\ 
 & u_{vbt}^{-1/2}\!\!=\!\!\left[ {{C}_{vs}}\!\left| {{S}_{Pb}} \right\rangle\! +\!\frac{{{C}_{vp}}}{\sqrt{3}}\!\left( \left| {{X}_{hal,X}} \right\rangle\! +\!\left| {{Y}_{hal,Y}} \right\rangle \!+\!\left| {{Z}_{hal,Z}} \right\rangle  \right) \right]\!\!\downarrow+\\&+u_{v,res}^{-1/2} \,.
\end{split}
\label{eq:eq10}
\end{equation}
%
%\cD{$S$ is not defined.}
Here $\left| {S} \right\rangle$ is the s-type atomic orbital, while $\left| {X} \right\rangle$, $\left| {Y} \right\rangle$, $\left| {Z} \right\rangle$ are the three p-type orbitals. The subscripts $\text{Pb}$, $hal,X$, $hal,Y$, and $hal,Z$ denote the atomic orbitals of lead and of the halogens having the nearest lead atoms along the $X$, $Y$, and $Z$ cubic axes, respectively.  $u_{c(v),res}^{+1/2}$ and $u_{c(v),res}^{-1/2}$ are the residual parts of the Bloch amplitudes, whose coupling with the nuclear spins is negligible. The coefficients $C_{vs}$, $C_{vp}$, $C_{cs}$, and $C_{cp}$ define the contributions of the $s$- and $p$-states to the vbt and cbb Bloch amplitudes. %According to \cite{boyer2016symmetry, targhi2018mapbi3}, $|C_{cs}|^2 \ll 1$.

The microscopic Hamiltonian of the hyperfine interaction \cite{abragam1961principles} reads:
\begin{gather}
    {{\widehat{H}}_{hf}}=-\hbar {{\gamma }_{N}}\hat{\vec{I}}\cdot {\widehat{\vec{B}}_{ae}}
    \label{eq:eq11},\\
\intertext{where}
    {{\widehat{\vec{B}}}_{ae}}=-2{{\mu }_{B}}\left[ \frac{{\vec{l}}}{{{r}^{3}}}-\frac{{\vec{s}}}{{{r}^{3}}}+3\frac{\vec{r}\left( \vec{s}\cdot \vec{r} \right)}{{{r}^{5}}}+\frac{8}{3}\pi \vec{s}\delta \left( {\vec{r}} \right) \right]
    \label{eq:eq12}
\end{gather}
is the magnetic field produced by the orbital angular momentum $\vec{l}$ and spin $\vec{s}$ of the atomic electron at the nucleus. Here $\vec{r}$ is the position vector of the atomic electron with respect to the nucleus.
%\cD{define $r$. Can we give name for ${\widehat{\vec{B}}}_{ae}$?}

The spin Hamiltonians for the interaction of the lead and halogen nuclei with the cbb electrons and the vbt holes are to be found by calculating the matrix elements of the operator $\widehat{\vec{B}}_{ae}$ on the wave functions of the corresponding charge carriers. The cbb and vbt spin Hamiltonians have the general form:
\begin{equation}
    \begin{split}
    & \widehat{H}_{i}^{cbb}={{\upsilon }_{0}}{{\left| \Psi _{e}\left( {{{\vec{R}}}_{i}} \right) \right|}^{2}}{{{\vec{I}}}_{i}}\cdot \hat{A}_{i}^{cbb}{{{\vec{S}}}_{e}}, \\ 
    & \widehat{H}_{i}^{vbt}={{\upsilon }_{0}}{{\left| \Psi _{h}\left( {{{\vec{R}}}_{i}} \right) \right|}^{2}}{{{\vec{I}}}_{i}}\cdot \hat{A}_{i}^{vbt}{{{\vec{S}}}_{h}}. \\ 
    \end{split}
    \label{eq:eq13}
\end{equation}
Here $\upsilon_0$  is the unit cell volume,  $\left| \Psi _{e}\left( {{{\vec{R}}}_{i}} \right) \right|^2$ and  $\left| \Psi _{h}\left( {{{\vec{R}}}_{i}} \right) \right|^2$ are the envelope functions of the cbb electron and the vbt hole, $i$ indexes the nuclei. The spins of the charge carriers are vector operators with components
\begin{equation}
    \hat{S}_{e}^{\alpha }\!=\!\frac{1}{2}\hat{\sigma }_{\alpha }^{e}, \,\, \;\hat{S}_{h}^{\alpha }\!=\!\frac{1}{2}\hat{\sigma }_{\alpha }^{h},
    \label{eq:eq14}
\end{equation}
where the Pauli matrices $\hat{\sigma }_{\alpha }^{e}$  and $\hat{\sigma }_{\alpha }^{h}$  are defined on the basis states given by Eqs.~\eqref{eq:eq9} and \eqref{eq:eq10}, respectively, $\alpha \in \{x,y,z\}$. The components of the hyperfine coupling tensors $\hat{A}_i^{cbb}$ and $\hat{A}_i^{vbt}$ are found as
\begin{equation}
    \begin{split}
    & \hat{A}_{i,\beta \alpha }^{cbb}=\hbar {{\gamma }_{Ni}}\operatorname{Tr}\left\{ \widehat{B}_{ae,\beta }^{i,cbb}\hat{\sigma }_{\alpha }^{e} \right\}, \\ 
    & \hat{A}_{i,\beta \alpha }^{vbt}=\hbar {{\gamma }_{Ni}}\operatorname{Tr}\left\{ \widehat{B}_{ae,\beta }^{i,vbt}\hat{\sigma }_{\alpha }^{h} \right\}, \\ 
    \end{split}
    \label{eq:eq15}
\end{equation}
where $\widehat{B}_{ae,\beta }^{i,cbb}$  and $\widehat{B}_{ae,\beta }^{i,vbt}$  are the matrices of the electron field operator at the $i$-th nucleus (the $\beta \in \{x,y,z\}$ denote the Cartesian components of the field), calculated with the conduction and valence band Bloch amplitudes, respectively.

Using the explicit forms of the $s$- and $p$-type atomic orbitals,
\begin{align*}
|S\rangle =& \psi_0(r),\\
|X\rangle =& \sqrt{\frac{3}{4\pi}}\psi_1(r)\frac{x}{r}=-\frac{1}{\sqrt{2}}\left(Y^{+1}_1-Y^{-1}_1\right)\psi_1(r),\\
|Y\rangle=& \!\sqrt{\frac{3}{4\pi}}\psi_1(r)\frac{y}{r}=i\frac{1}{\sqrt{2}}\left(Y^{+1}_1+Y^{-1}_1\right)\psi_1(r),\\
|Z\rangle =& \sqrt{\frac{3}{4\pi}}\psi_1(r)\frac{z}{r}=Y^0_1\psi_1(r),
\end{align*}
\begin{equation}
\label{eq:eq15a}
\end{equation}
one finds the hyperfine coupling tensors for the vbt holes and the cbb electrons. The interaction of the valence band holes with the lead nuclei is isotropic, and the corresponding hyperfine tensor is a scalar
\begin{equation}
    A_{Pb}^{vbt}=\frac{16}{3}\pi \hbar {{\gamma }_{Pb}}{{\mu }_{B}}{{\left| {{C}_{vs}} \right|}^{2}}{{\left| {{\psi }_{0,Pb}}\left( 0 \right) \right|}^{2}}=A_{Pb}^{s}{{\left| {{C}_{vs}} \right|}^{2}},
    \label{eq:eq16}
\end{equation}\\
where $A^s_{Pb}=\frac{16}{3}\pi \hbar {{\gamma }_{Pb}}{{\mu }_{B}}{{\left| {{\psi }_{0,Pb}}\left( 0 \right) \right|}^{2}}$. 
%\cD{Kirill, check as in equation ${\left| {\psi}_{0,Pb}(0) \right|}^{2}$ and below ${\left| {\psi}_{Pb}(0) \right|}^{2}$. Also, I do not see where we define $ {\psi}_{Pb}(0) $.} 
The interaction of the holes with the halogen nuclei is described by the tensors diagonal in the principal cubic axes. For the halogen atom having nearest lead atoms along the $Z$ axis it reads
\begin{equation}
\begin{split}
        \hat{A}_{hal,Z}^{vbt}=&\frac{4}{15}\hbar {{\gamma }_{hal}}{{\mu }_{B}}{{\left| {{C}_{vp}} \right|}^{2}}{{\left\langle \frac{1}{{{r}^{3}}} \right\rangle }_{p,hal}}\left( \begin{matrix}
            -1 & 0 & 0  \\
            0 & -1 & 0  \\
            0 & 0 & 2  \\
        \end{matrix} \right)=\\
        =&\frac{1}{20}A_{hal}^{p}{{\left| {{C}_{vp}} \right|}^{2}}\left( 
        \begin{matrix}
            -1 & 0 & 0  \\
            0 & -1 & 0  \\
            0 & 0 & 2  \\
        \end{matrix} \right),
\end{split}
\label{eq:eq17}
\end{equation}
where $$\left\langle\frac{1}{r^3}\right\rangle_p=\int\limits_{0}^{\infty}{\frac{1}{r^3}\psi_1(r)^2r^2dr},$$ and $$A^p_{hal}=\frac{16}{3}\hbar\gamma_{hal}\mu_B\left\langle \frac{1}{r^3}\right\rangle_p.$$ The hyperfine tensors for halogens having nearest lead atoms along $X$ and $Y$ are obtained from Eq.~\eqref{eq:eq17} by permutation of the Carthesian indices.  %\cD{which indices? There are several of them in (18).}

Similar results can be obtained for the interaction of the electrons with the lead and halogen nuclei. The isotropic interaction of the electron with the lead nuclei is characterized by the hyperfine constant:
\begin{equation}
    A_{Pb}^{cbb}=A_{Pb}^{p}{{\left| {{C}_{cp}} \right|}^{2}},
    \label{eq:eq18}
\end{equation}
where $A^{p}_{Pb}=\frac{16}{3}\hbar\gamma_{Pb}\mu_B\left\langle \frac{1}{r^3}\right\rangle_{p,Pb}$.
The anisotropic interaction of the electron and the halogen nuclei having nearest lead atoms along the $Z$ axis is described by a tensor diagonal in principal cubic axes:
\begin{equation}
    \hat{A}_{hal,Z}^{cbb}=\frac{{{\left| {{C}_{cs}} \right|}^{2}}}{3}A_{hal}^{s}\left( \begin{matrix}
   -1 & 0 & 0  \\
   0 & -1 & 0  \\
   0 & 0 & 1  \\
    \end{matrix} \right),
\label{eq:eq19}
\end{equation}
where $A^{s}_{hal}=\frac{16}{3}\pi\hbar\gamma_{hal}\mu_B\left|\psi_{0,hal}(0)\right|^2$. 
%\cD{define $\psi_{hal,Z}(0)$} 
The hyperfine tensors for the halogens having nearest lead atoms along $X$ and $Y$ are obtained from Eq. \eqref{eq:eq19} by permutation of the indices.

The values of $\left|\psi(0)\right|^2$ and $\left\langle1/r^3\right\rangle_p$ for the atomic orbitals of most chemical elements, calculated within the Hartree-Fock approximation, are tabulated in Refs. \cite{morton1978atomic, koh1985hyperfine}. Table \ref{tab:hyperfineparam} shows the values for the nuclear species relevant for the lead halide perovskites.

Concerning the hyperfine (hf) constants collected in Table \ref{tab:hyperfineparam} and their relevance to the lead halide perovskites, the following comments are in order: 

1) For the heavy atoms like lead and iodine, Hartree-Fock calculations give underestimated values of the hf constants. The relativistic theory still demonstrates a considerable disagreement with experiment, as best seen from the comparison of the theoretical (relativistic) and the experimental hf constants for the 5s shell of iodine (Ref. \cite{luc1975etude}). The empiric Mackey-Wood factor ($F_{MW}$) is believed to give the best approximation for the s-shell hf constants. However, in the case of iodine the discrepancy is still large. It is not clear whether the Mackey-Wood factor should be applied in calculating p-shell constants also. We could not find any discussion of this matter in literature. 

2) All the listed experimental constants, except those of Pb in PbTe~\cite{hewes1973nuclear}, were obtained from the optical spectra of atoms and ions. As seen from the data on PbTe, in semiconductor crystals the hf constants can differ from their atomic values. In particular, Hewes et al.~\cite{hewes1973nuclear} ascribe the smaller value of the s-shell hf constant of lead in PbTe, as compared to the atomic value reported by Schawlow et al.~\cite{schawlow1949hyperfine}, to the difference of the charge states of the lead ions between the two experiments. In PbTe, the effective charge of the Pb atoms is 1.3, while in atomic beam spectroscopy $\text{Pb}^{3+}$ ions were used. We would like to note that the valence band hole in lead halide perovskites corresponds to the absence of one of the two remaining s-electrons on the outer shell of the $\text{Pb}^{2+}$ ion, making the electron configuration similar to that of $\text{Pb}^{3+}$. Therefore, the atomic hf constant seems to be appropriate in our case. Also, Hewes et al.~\cite{hewes1973nuclear} reported an unusually large p-shell hf constant of lead in PbTe as compared to the theoretical atomic value. It is not clear how placement of the lead atom in a crystal can result in an almost 3-fold enhancement of the hyperfine coupling. Most likely, the theoretical p-shell constant of lead is underestimated. Its multiplication by the Mackey-Wood factor yields a value close to the one measured for PbTe, see Table \ref{tab:hyperfineparam}.

3) Recent experimental studies on carrier spin precession in the hyperfine fields of the nuclear spin fluctuations in perovskite nanocrystals~\cite{Meliakov2024,Meliakov2025} give experimental estimates of the hyperfine constants in lead halide perovskites. In particular, the average dispersion $\Delta_{E,h}=4$~$\mu$eV of the hole spin splitting in the nuclear spin fluctuations in CsPbBr$_{3}$ nanocrystals with the average diameter of 9~nm \cite{Meliakov2024} corresponds to $A_{Pb}^{vbt}\approx280$~$\mu$eV (see Eq.~(A11) in Ref.~\onlinecite{Meliakov2024}). Similarly, the dispersion of the electron spin splitting of 0.8~$\mu$eV in CsPbI$_3$ nanocrystals corresponds to ${\left| {{C}_{cs}} \right|}^{2}A_{I}^{s}\approx17$~$\mu$eV \cite{Meliakov2025}. With ${\left| {{C}_{cs}} \right|}^{2}=0.09$ obtained by DFT calculations, $A_{I}^{s}\approx190$~$\mu$eV was estimated, which is 30\% less than the value experimentally determined earlier by atomic spectroscopy \cite{luc1975etude}.  

\subsection{Dynamic polarization of nuclear spins by electrons and holes}
\label{subsec:DNP theory}

\subsubsection{Dynamic polarization in strong external magnetic fields: Symmetry considerations}
\label{sec:theory_DNP_strong_fields}

The hyperfine interaction with spin-polarized charge carriers results in a repopulation of the nuclear spin states, known as the dynamic polarization of the nuclear spins \cite{abragam1961principles, OOChapter5}. If an applied external magnetic field is much stronger than the local fields of the nuclear spin-spin interactions, the eigenstates of the nuclear spins are determined by the Zeeman interaction with this field as well as by the quadrupole splitting (in our case of the halogen nuclei). The polarized nuclear spins create the effective magnetic fields, which split the spin states of the charge carriers. These effective fields, mediated by the hyperfine interaction, are called Overhauser fields. Experimentally they can be measured via the Hanle and polarization recovery effects.

In the lead halide perovskites there are several nuclear species, some of which experiencing a strong quadrupole splitting and an anisotropic hyperfine coupling with the charge carriers. For this reason, in order to analyze the pattern of the Overhauser fields created by the dynamic nuclear polarization, we first apply symmetry considerations. In the limit of weak spin polarization, the Cartesian components of the Overhauser field are linear functions of the components of the carrier mean spin $\left\langle \vec{S}\right\rangle$, defined by a second-rank tensor:
\begin{equation}
    B_{N, \alpha}=\sum\limits_{\beta}{a_{\alpha\beta}\left\langle S\right\rangle_\beta}.
    \label{eq:eq20}
\end{equation}
%
%\cD{define $\left\langle S\right\rangle$, what is $\alpha$ and $\beta$? Is $\alpha=x,y,z$?}
where $B_{N, \alpha}$ and $\left\langle S\right\rangle_\beta$ represent Carthesian components of the vectors $\vec{B}_N$ and $\left\langle \vec{S}\right\rangle$. 
The tensor components $a_{\alpha\beta}$ are functions of the external magnetic field, which are symmetric with respect to field inversion. Due to the overall cubic symmetry of the crystal, the $a_{\alpha\beta}$ satisfy the following conditions \cite{OOChapter5}:
\begin{widetext}
\begin{equation}
    \begin{split}
        a_{xx}\left(B_x,B_y,B_z\right)&= \;a_{xx}\left(B_x,B_z,B_y\right)\\
        a_{xy}\left(B_x,B_y,B_z\right)&= \;a_{yx}\left(B_y,B_x,B_z\right)=
        -a_{xy}\left(-B_x,B_y,B_z\right)=-a_{xy}\left(B_x,-B_y,B_z\right)=a_{xy}\left(B_x,B_y,-B_z\right),
    \end{split}
    \label{eq:eq21}
\end{equation}
\end{widetext}
with similar relations for the other components obtainable by permutation of the indices. Since the nuclear spin polarization for $B \gg B_L$ is virtually independent of the field strength $B$, $a_{\alpha\beta}$ can be considered as functions of the direction cosines of the external field, $b_x=B_x/B$, $b_y=B_y/B$, and $b_z=B_z/B$. Expanding $a_{\alpha\beta}$  in powers of $b_x$, $b_y$ and $b_z$, we obtain:
\begin{widetext}
\begin{equation}
    \begin{split}
         & B_{N,x}=a_{11}b_x^2\left\langle S_x \right\rangle +a_{12}\left( b_y^2+b_z^2\right)\left\langle S_x \right\rangle +a_{44}\left( b_xb_y\left\langle S_y \right\rangle +b_xb_z\left\langle S_z \right\rangle  \right)+\delta B_{N,x}^{(4)}, \\ 
         & B_{N,y}=a_{11}b_y^2\left\langle S_y \right\rangle +a_{12}\left( b_x^2+b_z^2 \right)\left\langle S_y \right\rangle +a_{44}\left( b_yb_x\left\langle S_x \right\rangle +b_yb_z\left\langle S_z \right\rangle  \right)+\delta B_{N,y}^{(4)}, \\ 
        & B_{N,z}=a_{11}b_z^2\left\langle S_z \right\rangle +a_{12}\left( b_x^2+b_y^2\right)\left\langle S_z \right\rangle +a_{44}\left( b_zb_y\left\langle S_y \right\rangle +b_zb_x\left\langle S_x \right\rangle  \right)+\delta B_{N,z}^{(4)}, \\ 
    \end{split}
\label{eq:eq22}
\end{equation}
\end{widetext}
where $a_{11}=a_{\alpha\alpha\alpha\alpha}$, $a_{12}=a_{\alpha\alpha\beta\beta}$, and $a_{44}=a_{\alpha\beta\alpha\beta}$ are the three independent components of the fourth-rank tensor $\hat{a}$, allowed by the cubic symmetry, and $\delta B^{(4)}_{N,x}$, $\delta B^{(4)}_{N,y}$, $\delta B^{(4)}_{N,z}$ are corrections of 4-th power in the external field components. To comply with the symmetry relations given by Eq.~\eqref{eq:eq21}, these corrections should have the form
\begin{widetext}
\begin{equation}
    \begin{split}
    & \delta B_{N,x}^{(4)}={{C}_{1}}b_{x}^{4}\left\langle {{S}_{x}} \right\rangle +{{C}_{2}}\left( b_{y}^{4}+b_{z}^{4} \right)\left\langle {{S}_{x}} \right\rangle +{{C}_{3}}\left( {{b}_{x}}b_{y}^{3}\left\langle {{S}_{y}} \right\rangle +{{b}_{x}}b_{z}^{3}\left\langle {{S}_{z}} \right\rangle  \right)+{{C}_{4}}\left( {{b}_{y}}b_{x}^{3}\left\langle {{S}_{y}} \right\rangle +{{b}_{z}}b_{x}^{3}\left\langle {{S}_{z}} \right\rangle  \right), \\ 
    & \delta B_{N,y}^{(4)}={{C}_{1}}b_{y}^{4}\left\langle {{S}_{y}} \right\rangle +{{C}_{2}}\left( b_{x}^{4}+b_{z}^{4} \right)\left\langle {{S}_{y}} \right\rangle +{{C}_{3}}\left( {{b}_{y}}b_{x}^{3}\left\langle {{S}_{x}} \right\rangle +{{b}_{y}}b_{z}^{3}\left\langle {{S}_{z}} \right\rangle  \right)+{{C}_{4}}\left( {{b}_{x}}b_{y}^{3}\left\langle {{S}_{x}} \right\rangle +{{b}_{z}}b_{y}^{3}\left\langle {{S}_{z}} \right\rangle  \right), \\ 
    & \delta B_{N,z}^{(4)}={{C}_{1}}b_{z}^{4}\left\langle {{S}_{z}} \right\rangle +{{C}_{2}}\left( b_{y}^{4}+b_{x}^{4} \right)\left\langle {{S}_{z}} \right\rangle +{{C}_{3}}\left( {{b}_{z}}b_{y}^{3}\left\langle {{S}_{y}} \right\rangle +{{b}_{z}}b_{x}^{3}\left\langle {{S}_{x}} \right\rangle  \right)+{{C}_{4}}\left( {{b}_{y}}b_{z}^{3}\left\langle {{S}_{y}} \right\rangle +{{b}_{x}}b_{z}^{3}\left\langle {{S}_{x}} \right\rangle  \right), \\ 
    \end{split}
    \label{eq:eq23}
\end{equation}
\end{widetext}
where $C_1$, $C_2$, $C_3$, and $C_4$ are independent constants.

Now we turn to the calculation of the contributions of specific spin states to the Overhauser field. In case of $^{207}\text{Pb}$, the eigenstates of the individual nuclear spins in the external field are the states with the definite projections of the nuclear spin on the external field, $I_B=\pm 1/2$. Since the hyperfine interaction of both electrons and holes with the Pb nuclei is isotropic, the dynamic polarization results from the flip-flop transitions conserving the total spin of the carrier and the nucleus. In the absence of spin leakage, an equilibrium establishes with projections of the mean spins of nuclei and carriers equal to each other: $\left\langle I_B\right\rangle=\left\langle S_B\right\rangle$, where $S_B$ is the spin projection of the corresponding charge carrier on the magnetic field. The spin polarization of $^{207}\text{Pb}$ nuclei results in the appearance of an Overhauser field, which is parallel or antiparallel to the external field and, according to Eqs.~\eqref{eq:eq16} and \eqref{eq:eq18}, for electrons equals
\begin{align}
    {{B}^{{Pb}}_{N,e}}=\frac{A_{Pb}^{cbb}{{x}_{Pb}}\left\langle {{S}_{e,B}} \right\rangle }{{{\mu }_{B}}{{g}_{e}}}	 \label{eq:eq24} \,.\\
\intertext{For holes it equals}
    {{B}^{{Pb}}_{N,h}}=\frac{A_{Pb}^{vbt}{{x}_{Pb}}\left\langle {{S}_{h,B}} \right\rangle }{{{\mu }_{B}}{{g}_{h}}} \,.
    \label{eq:eq25}
\end{align}
Here $\mu_B$ is the Bohr magneton, $g_e$  and $g_h$ are the $g$-factors of electrons and holes, and $x_{{Pb}}$ is the abundance of the magnetic isotope $^{207}\text{Pb}$. The Overhauser field adds up to the external field or subtracts from it, depending on the sign of the carrier $g$-factor and the mutual orientation of the carrier spin and the magnetic field. The total field acting on the carrier spin remains collinear to the external field. This corresponds to $a_{12}=0$ and zero fourth-power corrections, while $a_{11}\!=\!a_{44}\!=\!A^{cbb}_{Pb}x_{Pb}/\left(\mu_Bg_e\right)$ for the electrons and $a_{11}\!=\!a_{44}\!=\!A^{vbt}_{Pb}x_{Pb}/\left(\mu_Bg_h\right)$ for the holes.

The pattern of the dynamic spin polarization of the halogen nuclei is much more complex due to their quadrupole splitting and anisotropic hyperfine coupling. For the $M_Q=\pm 3/2$ and $M_Q=\pm 5/2$ doublets, having extremely anisotropic $g$-factors, the quantization axis coincides with the quadrupole axis ($X$, $Y$, or $Z$ depending on the nucleus position with respect to the nearest Pb atom), except the special case of the external field directed exactly perpendicular to that axis \cite{OOChapter5, artemova1985polarization}. Applying the principle of detailed balance to transitions between these doublets and between the $M_Q=\pm 1/2$ and $M_Q=\pm 3/2$ doublets, induced by the charge carrier, and conserving the total spin projection on the quadrupole axis, one arrives to the following equilibrium value of the mean spin associated with the $M_Q=\pm 3/2$ and $M_Q=\pm 5/2$ doublets:
\begin{align}
    \left\langle {{I}_{Q}} \right\rangle =\frac{8}{3}{{\left\langle {{I}_{1/2}} \right\rangle }_{Q}}+\frac{26}{3}{{\left\langle S \right\rangle }_{Q}} \label{eq:eq26} \,, \\
    \intertext{for $I=5/2$ (iodine), and}
    \left\langle {{I}_{Q}} \right\rangle =\frac{3}{2}{{\left\langle {{I}_{1/2}} \right\rangle }_{Q}}+3{{\left\langle S \right\rangle }_{Q}} \,,
    \label{eq:eq27}
\end{align}
for $I=3/2$ (bromine and clorine). Here $\left\langle S\right\rangle_Q$ is the carrier spin projection on the quadrupole axis. $\left\langle I_{1/2}\right\rangle_Q \leq \left\langle S\right\rangle_Q$ is the projection of the mean spin of the $M_Q=\pm 1/2$ doublet on the same axis, which depends not only on the carrier spin, but also on the external field. One can separate the external field-independent part of the mean spin of the quadrupole doublets $I_{QS}$, that is proportional to $\left\langle S\right\rangle_Q$, from the field-dependent part $I_{QB}$. The associated total Overhauser field including the contributions of the quadrupole doublets with all orientations ($X$, $Y$, $Z$) is collinear with the carrier mean spin, and for this reason, it does not contribute to the Hanle effect \cite{OOChapter5}. However, it does contribute to the polarization recovery in the longitudinal magnetic field. For example, in the case of iodine ($I=5/2$), for the conduction band this field equals to
%\cD{Kirill, seems that we can skip "In case of iodine", as we generalize story for "halogens". Means also that we should probably use in equations below ${{\vec{B}}^{I}_{NQSc}}$ as ${{\vec{B}}^{hal}_{NQSc}}$. Or you plan turn here already to iodine?} 
%
\begin{align}
    {{\vec{B}}^{I}_{NQSc}}=\frac{26{{\left| {{C}_{cs}} \right|}^{2}}A_{hal}^{s}}{9{{\mu }_{B}}{{g}_{e}}}\left\langle {{{\vec{S}}}_{e}} \right\rangle \label{eq:eq28} \,.\\
    \intertext{For the valence band, it is equal to}
    {{\vec{B}}^{I}_{NQSv}}=\frac{13{{\left| {{C}_{vp}} \right|}^{2}}A_{hal}^{p}}{15{{\mu }_{B}}{{g}_{h}}}\left\langle {{{\vec{S}}}_{h}} \right\rangle \,.
    \label{eq:eq29}
\end{align}
This corresponds to zero $a_{44}$ and fourth-power corrections, while $a_{11}=a_{12}=26\left|C_{cs}\right|^2A^s_{hal}/\left(9\mu_Bg_e\right)$ for the electrons and $a_{11}=a_{12}=13\left|C_{vp}\right|^2A^p_{hal}/\left(15\mu_Bg_h\right)$ for the holes.

The rather cumbersome evaluation of the hyperfine field resulting from the dynamic polarization of $M_q=\pm 1/2$, presented in detail in Subsec.~\ref{sec:DNP}, yields the coefficients for Eqs.~\eqref{eq:eq22} and \eqref{eq:eq23}. They connect the components of the Overhauser field with the direction cosines of the external field. These coefficients for the contributions from lead and all the halogens are listed in Tables \ref{tab:coeffCondband} and \ref{tab:coeffValenceband} for the conduction and valence bands, respectively.

Using these coefficients, one can calculate the Overhauser field for any specific geometry. For instance, in the polarization recovery geometry (both the mean spin and the external field parallel to $Z$) for the case $Z \;\| \left[001\right]$
\begin{equation}
    {{B}_{Nz}}=\left( {{a}_{11}}+{{C}_{1}} \right)\left\langle {{S}_{z}} \right\rangle. 
    \label{eq:eq30}
\end{equation}
In the oblique magnetic field used for the Hanle effect measurements the dependence is much more complicated. However, one can use a simplified expression to calculate the strength of the external field needed to compensate the Overhauser field (e.g., the position of the side maximum on the Hanle curve). Indeed, under these conditions, the perpendicular component of the total field applied to the carrier spin is close to zero, and for this reason, the mean spin direction is approximately parallel to the wave vector of the excitation light. If the latter is parallel to $Z$, the nuclear field at the point of compensation equals
\begin{multline}
    {{B}_{Ncomp}}=\big[ {{a}_{44}}\cos \theta +{{C}_{3}}{{\cos }^{3}}\theta +\\+{{C}_{4}}\cos \theta {{\sin }^{2}}\theta \left( {{\cos }^{4}}\varphi +{{\sin }^{4}}\varphi  \right) \big]\left\langle {{S}_{z}} \right\rangle. 
    \label{eq:eq31}
\end{multline}
The contributions from different nuclei are additive, so that $B_{N, e(h)}=B^{{Pb}}_{N, e(h)} + B^{{hal}}_{N, e(h)}$.  
%\cD{Check $B^{{I}}_{N, e(h)}$ versus $B^{{hal}}_{N, e(h)}$?}

\subsubsection{Dynamic polarization of quadrupole-split nuclei via anisotropic hyperfine interaction: finding the coefficients}
\label{sec:DNP}

The hyperfine interaction leads to mutual spin flips of the nucleus and the charge carrier. If the electron or hole spin polarization is continuously generated via optical pumping, this will polarize the nuclei. In the case of an isotropic hyperfine interaction, only flip-flop transitions are allowed, which conserve the total spin of the electron-nuclear spin system. The presence of anisotropy makes flip-flip/flop-flop transitions possible, in which both the nuclear and electron spins increase or decrease by one.

Let us consider the problem of dynamic polarization of quadrupole-split nuclei by electrons/holes with an average spin $\langle S\rangle$ in the presence of a magnetic field $B$, which is much larger than the local field. Let the $Z$-axis be directed along the principal quadrupole axis. Then, the directions of $\langle S\rangle$ and $B$ will be defined by the angles $\omega, \chi$ and $\theta, \varphi$ in the spherical coordinate system. In the general case, the transition rate from nuclear state $n$ to $m$ is given by the squared modulus of the matrix element of the hyperfine interaction operator between these states, averaged over the electronic states:
\begin{equation}
\begin{split}
    W_{n\rightarrow m} &= \sum\limits_{p,q}{n_n^{e(h)}\left| \left\langle  \phi_mu_p \right|\hat{H}_{hf}\left| \phi _nu _q \right\rangle  \right|^2} \,,\\
    \psi^{e(h)}_q &= \sum\limits_q{D_{q,p}\left(0,\omega,\chi\right)u_p.}
\end{split}
\label{eqSM:eq1}
\end{equation}
Here, $\hat{H}_{hf}$ is the Hamiltonian of the hyperfine interactions for electrons or holes defined in Eq. \eqref{eq:eq13}; $\psi^{e(h)}_q$ are the eigenfunctions of the $S_z$ operator;  $D_{q,p}\left(0,\omega,\chi\right)$ is the finite rotation matrix; $n_{1/2}^{e(h)}$ and  $n_{-1/2}^{e(h)}$ are the populations of the corresponding spin states of charge carriers. Their ratio is determined by the average polarization of electrons (holes): $\frac{n_{1/2}^{e(h)}}{n_{-1/2}^{e(h)}} = \frac{1+2\langle S\rangle}{1-2\langle S\rangle}$.

The spin states of the halogen nuclei are eigenfunctions of the following Hamiltonian:
\begin{equation}
        {{\hat{H}}_{Q}}={{A}_{Q}}\left( 3\hat{I}_{Z}^{2}-I\left( I+1 \right) \right) -\hbar\gamma_{hal}\left(\hat{\vec{I}}\cdot\vec{B}\right),
        \label{eqSM:eq2}
\end{equation}
where $Z$ is the EFG axis. The first term corresponds to the quadrupole interaction (we neglect the weak non-axiality), and the second term represents the Zeeman interaction. In weak magnetic fields (up to 100 mT) used in the experiments, the second term can be treated as a perturbation. To first order, we obtain the following set of eigenfunctions:
\begin{equation}
    \begin{split}
        |M_Q| > 1/2 \hspace{6pt}&\phi_{M_Q} = \psi_{M_Q} \,, \\
        M_Q = 1/2 \hspace{6pt}&\phi_{+} = \sin(\alpha)\psi_{1/2}+e^{i\varphi}\cos(\alpha)\psi_{-1/2} \,,\\
        M_Q = -1/2 \hspace{6pt}&\phi_{-} = -e^{-i\varphi}\cos(\alpha)\psi_{1/2}+\sin(\alpha)\psi_{-1/2} \,,
    \end{split}
    \label{eqSM:eq3}
\end{equation}
where 
\begin{equation*}
\begin{split}
\tan{\alpha} = 
 \begin{cases}
   \sqrt{\frac{f+1}{f-1}} & \cos{\theta} \ge 0\\
   \sqrt{\frac{f-1}{f+1}} & \cos{\theta} < 0
 \end{cases}
 \\
 f = \sqrt{1+\left(I+1/2\right)^2\tan{\theta}}.
 \end{split}
\end{equation*}
The strong quadrupole interaction fixes the eigenfunctions of the nuclear states with spin $|M_Q| > 1/2$. Regardless of the external magnetic field direction, these states remain eigenfunctions of the $\hat{I}_Z$ operator. For the doublet states with $M_Q = \pm 1/2$, a complex angular dependence on the magnetic field direction is observed.

To calculate the nuclear polarization induced by polarized charge carriers, we solved the stationary system of population balance equations for the nuclear levels:
\begin{equation}
    -\sum\limits_{m}{{n}_{n}}W_{n\rightarrow m}+\sum\limits_{n\ne m}{{{n}_{m}}W_{m\rightarrow n}}=0.
    \label{eqSM:eq4}
\end{equation}
The resulting population values enable calculation of the average nuclear spin and the Overhauser field. The total Overhauser field from the halogen nuclei  includes distinct contributions from the nuclei located along the hal-Pb-hal chains oriented along different cubic axes ([100], [010], [001]).

In the case of weak carrier polarization, the Overhauser field can be expanded into a series in terms of the average carrier spin. If one keeps only the linear term, then the resulting expression can be fitted using Eqs. \eqref{eq:eq22} and \eqref{eq:eq23}. For this purpose, several orientations of the external magnetic field were selected, and for each of these values, the Overhauser field was calculated for an arbitrary orientation of the average carrier spin. The expansion coefficients (Eqs.~\eqref{eq:eq22}, \eqref{eq:eq23}) were determined by minimizing the residual $Er$: 
\begin{multline}
    Er=\sum\limits_{p,q}{\big[ \vec{B}_N(\theta_p,\phi_q)} \\{- \vec{B}_N\left(a_{11},a_{12},a_{44},C_1,C_2,C_3,C_4,\theta_p,\phi_q\right)\big]^2}.
    \label{eqSM:eq5}
\end{multline}
The values of these coefficients are presented in Tables \ref{tab:coeffCondband} and \ref{tab:coeffValenceband}. Here, the first term represents the field obtained from the calculations, while the second term corresponds to the field calculated using Eqs.~\eqref{eq:eq22} and \eqref{eq:eq23}. The residual values at the minimum were approximately 0.1 and 0.01 mT$^2$ for the electrons with the nuclear spins $I=5/2$ and $I=3/2$, respectively, and 0.5 and 0.1 mT$^2$ for the holes with the nuclear spins $I=5/2$ and $I=3/2$, respectively. The resulting errors constitute roughly one thousandth of the sum $\sum\limits_{p,q}{{B}^2_N(\theta_p,\phi_q)}$.

\subsubsection{Dynamic polarization in weak external magnetic fields}
\label{sec:theory_DNP_weak_fields}

In external fields weaker than or comparable to the internal fields of the spin-spin interactions $B_L$ (see Subsec.~\ref{sec:dipole-dipoleInteraction}), the dynamic polarization is usually described in terms of nuclear spin cooling, i.e. the reduction of the absolute value of the spin temperature of the nuclear spin system (NSS)~\cite{OOChapter2, OOChapter5, SPSDyakonov}. The main point here is that the spin-spin interactions rapidly destroy the non-equilibrium spin polarization of the nuclei, but conserve the total energy of the NSS. Therefore, the energy of the Zeeman interaction, injected into the NSS when the nuclear magnetic moments are dynamically polarized in the presence of the external field, is redistributed between the Zeeman and the spin-spin energy reservoirs, reducing the NSS entropy. As a result, the spin temperature becomes lower than the lattice temperature in absolute value, while its sign is determined by the mutual orientation of the nuclear magnetization and the external field. The difference between the temperatures of the NSS and of the crystal lattice persists as long as the spin-lattice relaxation time, typically being in the range of $0.1-1000$ seconds \cite{vladimirova2017nuclear, kotur2018spin, gribakin2024nuclear}. While the spin temperature is low, the NSS demonstrates a greatly enhanced paramagnetic susceptibility, which results in the appearance of nuclear polarization parallel or antiparallel to the external field, depending on the spin temperature sign.

The value of the nuclear spin polarization as function of the applied field is described by the following expression \cite{Dyakonov1975,smirnov2025cooling}: 
\begin{equation}
    {{p}_{N}}=\frac{4\left( I+1 \right){{B}^{2}}\left\langle {{S}_{B}} \right\rangle }{3\left( {{B}^{2}}+\tilde{B}_{L}^{2} \right)}\,.
    \label{eq:eq32}
\end{equation}
If the carrier spin correlation time (i.e. the time of carrier hopping between localization centers) is short as compared to the time of carrier spin precession in the nuclear field, than $\tilde{B}_L^2 \approx 3B_L^2$  \cite{Dyakonov1975}. For a long carrier spin correlation time (the situation typical for perovskites \cite{kudlacik2024optical}), $\tilde{B}_L^2$ can be much larger than  $B_L^2$ \cite{smirnov2025cooling}. 

However, the application of the spin temperature theory to lead halide perovskites is not straightforward. First of all, the field-independent DNP of the quadrupole-split doublets (Eqs.~\eqref{eq:eq28} and \eqref{eq:eq29}) is expected to remain even at zero field, similarly to observations in GaAs quantum dots \cite{dzhioev2007stabilization}. Then, the local field at the Pb nuclei is mainly contributed not by their dipole-dipole interactions between each other, but by their interactions with the quadrupole-split halogen nuclei (see Subsec.~\ref{sec:dipole-dipoleInteraction}). This means that the theory of Ref.~\onlinecite{smirnov2025cooling} cannot be immediately applied here. Finally, the DNP of $M_Q=\pm 1/2$ doublets of the halogen nuclei is expected to be anisotropic. The theory of the weak-field DNP under these conditions remains to be developed. 

To summarize, one can expect two contributions to the DNP and the resulting Overhauser fields at weak external magnetic fields: a field-independent part due to polarization of the quadrupole-split halogen nuclei along the charge carrier spin, and a quadratic in magnetic field contribution similar to Eq.~\eqref{eq:eq32}. Notably, the magnetic field may include the internal field created by polarized quadrupole-split halogen nuclei (Eqs.~(\ref{eq:eq6}-\ref{eq:eq8})). This field would manifest itself as a shift of the peak on the Hanle curve or of the dip in the PR curve.

%%%%%%%%%%%%%%%%%%%%%%%%%%%%%%%%%%%%%%%%%%%%%%%%%%%%%%%%%%%

\section{Samples and experimental setup}
\label{sec:experimentals}

We study a bulk FA$_{0.9}$Cs$_{0.1}$PbI$_{2.8}$Br$_{0.2}$ crystal of lead halide perovskite, grown by the inverse temperature crystallization technique~\cite{nazarenko2017single}. Details of the synthesis procedure and optical properties including the carrier spin dynamics studied by time-resolved Kerr rotation can be found in Refs.~\cite{kirstein2022lead,kopteva2023giant,grisard2023long}.  
%This material has an inverted band structure, compared to the conventional III-V and II-VI semiconductors, which is characteristic to the lead halide perovskite semiconductors. 

The sample was placed in a liquid helium bath cryostat with a variable temperature insert, the temperature was tuned from $T=1.6$~K  up to 15~K. An external magnetic field, $\mathbf{B}$, with a strength up to 70~mT was provided by an electromagnet, which could be rotated in order to apply the field in Voigt geometry (field direction perpendicular to the light $k$-vector, $\mathbf{B_{\rm V} \perp \mathbf{k}}$), in Faraday geometry ($ \mathbf{B_{\rm F}} \parallel \mathbf{k}$) or in a tilted geometry with an angle $\theta$ between $\mathbf{k}$ and $\mathbf{B}$. The photoluminescence (PL) was excited with circularly polarized light emitted by a continuous wave (CW) Titanium-Sapphire laser, whose photon energy was set to $E_{\rm exc} = 1.590$~eV. The pump polarization was either modulated between $\sigma^+$ and $\sigma^-$ by a photoelastic modulator (PEM) operating at the frequency of 50 kHz, or kept fixed $\sigma^+$ or $\sigma^-$. The laser beam was directed along the [001] axis, while the photoluminescence was collected in the reflection geometry, spectrally dispersed by a 0.5-meter spectrometer, and detected using an avalanche photodiode. A two-channel photon counting device synchronized with the PEM enabled simultaneous measurement of the $I^{++}$ and $I^{-+}$ intensities of the $\sigma^+$ photoluminescence component under modulated $\sigma^+ / \sigma^-$ pumping (PEM in excitation). On this basis, the optical orientation degree, $P_\mathrm{oo}$, can be measured:
\begin{equation}
P_{\rm oo}=\frac{I^{++}-I^{-+}}{I^{++}+I^{-+}}.
\end{equation}
Alternatively, for constant $\sigma^+$ polarized excitation (PEM in detection), the optical orientation degree is given by:
\begin{equation}
P_{\rm oo}=\frac{I^{++}-I^{+-}}{I^{++}+I^{+-}},
\end{equation}
where $I^{++}$ and $I^{+-}$ denote the intensities of the photoluminescence in $\sigma^+$ and $\sigma^-$ polarization, respectively.

\begin{figure}[t!]
\center{\includegraphics{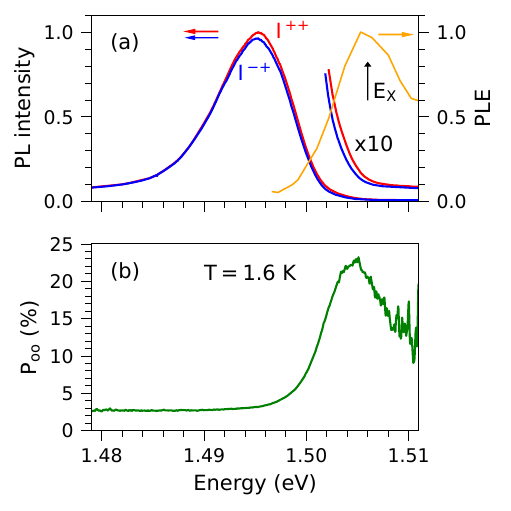}}
\caption{(a) Measured intensities $I^{++}$ (red line) and $I^{-+}$ (blue line) of the $\sigma^+$ component of the photoluminescence (PL) under $\sigma^+/\sigma^-$ excitation modulated at 50~kHz in a FA$_{0.9}$Cs$_{0.1}$PbI$_{2.8}$Br$_{0.2}$ crystal. $T=1.6$~K,  the excitation photon energy is $E_{\rm exc}=1.590$~eV and the excitation power density is $P=18$~ W/cm\textsuperscript{2}. Photoluminescence excitation spectra (PLE, orange line) measured at the detection energy of $E_{\rm det}=1.495$~eV set to the PL maximum. (b) Spectral dependence of the optical orientation degree.} \label{fig:spectrum}
\end{figure}

PL and photoluminescence excitation (PLE) spectra of FA$_{0.9}$Cs$_{0.1}$PbI$_{2.8}$Br$_{0.2}$ are shown in Fig.~\ref{fig:spectrum}(a). The PL has a maximum at 1.495~eV and, as was shown previously, it originates from the recombination of spatially separated localized carriers (electrons and holes) with additional exciton emission at higher energies~\cite{kirstein2022lead, kopteva2023giant, kudlacik2024optical}. The spectral dependence of the optical orientation degree, $P_{\rm oo}$, is shown in Fig.~\ref{fig:spectrum}(b). It has a maximum at 1.504~eV, which is close to the exciton resonance at $E_{\rm X} = 1.506$~eV observed in the PLE spectrum.  The exciton binding energy in the studied FA$_{0.9}$Cs$_{0.1}$PbI$_{2.8}$Br$_{0.2}$ sample should be close to the 14~meV known for FAPbI$_3$~\cite{galkowski2016determination}. This enables us to confirm the band gap energy value $E_{\rm g} = 1.520$~eV. The measured $P_{\rm oo}=23\%$ is contributed by electrons and holes as well as excitons due to the overlap of the emission of localized carriers and excitons~\cite{kudlacik2024optical}. It is remarkable, that this high degree of optical orientation is observed for a laser excitation energy which exceeds by 70 meV the band gap energy and by 84~meV the exciton energy. It was shown recently, that this is typical for the lead halide perovskite semiconductors, for which the energy relaxation of excitons and carriers is not accompanied by their spin relaxation, see Refs.~\onlinecite{kopteva2023giant,Kopteva_2025OOX,Kopteva_2025OOmapi,Zhukov_2025two_colors}. If not stated otherwise, the energy of 1.504~eV, at which the degree of optical orientation reaches its maximum, is selected as the detection energy $E_{\rm det}$ for the subsequent measurements.

\section{Experimental results}
\label{sec:exp}

\subsection{Spectral dependence of the Hanle and polarization recovery curves}
\label{sec:exp_spectral}

\begin{figure*}[t!]
\center{\includegraphics{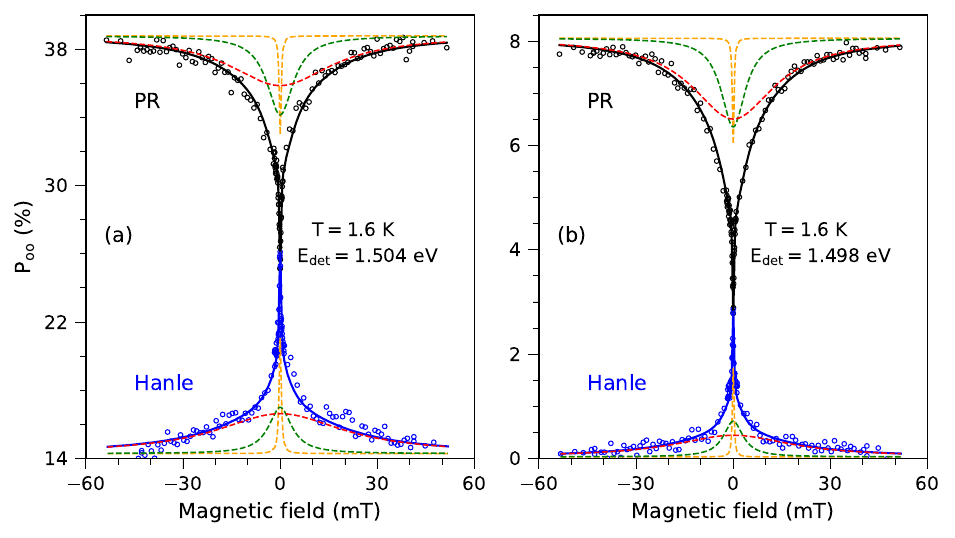}}
\caption{Hanle (blue symbols) and polarization recovery (black symbols) curves measured in Voigt and Faraday geometry, respectively, at the detection energies of: (a) $E_{\rm det}=1.504$~eV and (b) $E_{\rm det}=1.498$~eV. The excitation is $\sigma^+/\sigma^-$ modulated at the frequency of 50~kHz, $E_{\rm exc}=1.590$~eV, $P=18$~W/cm\textsuperscript{2}. The solid lines represent fits with three Lorentz functions, each shown individually by the dashed lines in red, green, and orange.} 
\label{fig:HPspectral}
\end{figure*}

\begin{figure*}
\center{\includegraphics{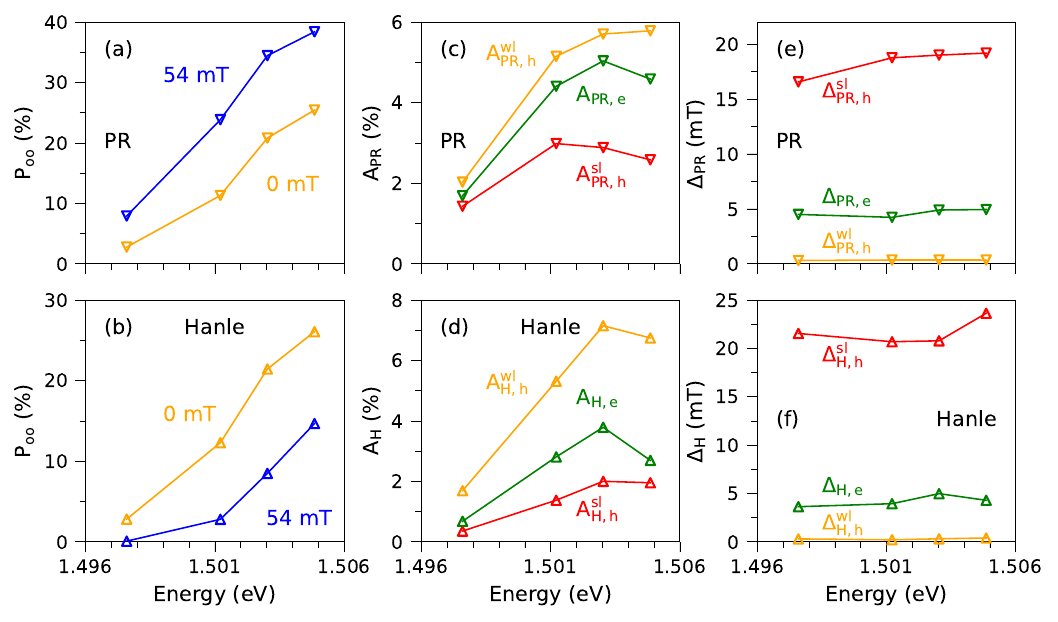}}
\caption{Spectral dependence of the parameters evaluated from the Hanle and polarization recovery curves. Excitation is $\sigma^+/\sigma^-$ modulated at the frequency of 50~kHz, $E_{\rm exc}=1.590$~eV, $P=18$~W/cm\textsuperscript{2}. (a,b) $P_{\rm oo}$ from the PR and Hanle curves measured at zero field (orange symbols) and 54~mT (blue symbols). (c,d) PR and Hanle amplitudes.  (e,f) PR and Hanle half-widths at half maximum. In panels (c-f) the results for the three fit components are shown, which correspond to strongly localized holes (red symbols), electrons (green symbols), and weakly localized holes (orange symbols). Lines in all panels are guides to the eye.}
\label{fig:HPspectral_parameters}
\end{figure*}

The phenomenon of luminescence depolarization in a transverse magnetic field (Voigt geometry) is known as the Hanle effect~\cite{meier2012optical}. Under CW excitation, when the external magnetic field is applied, the carrier spins start to precess around the field direction. This Larmor precession causes a decrease in the carrier mean spin polarization along the $k$-vector direction,  which determines the optical orientation degree $P_{\rm oo}(B_{\rm V})$. The Hanle curves measured at the exciton resonance energy of 1.504~eV and at the PL maximum of 1.498~eV  are shown by the blue circles  in Figs.~\ref{fig:HPspectral}(a) and \ref{fig:HPspectral}(b), respectively. The polarization of the laser beam is $\sigma^+/\sigma^-$ modulated at the frequency of 50~Hz. The Hanle curves have a complex shape and for fitting them we use a function consisting of three Lorentzian contours with different amplitudes and half-widths at half-maximum (HWHM), $\Delta_{\rm H}$. The fit components are shown by the dashed lines. The evaluated parameters and their spectral dependences are given in Figs.~\ref{fig:HPspectral_parameters}(b,d,f). The Hanle widths $\Delta_{\rm H}$ are about 20, 4, and 0.3~mT. It is clearly seen that with the change of detection energy, the widths $\Delta_{\rm H}$ of the three Lorentzian contours remain practically the same (Fig.~\ref{fig:HPspectral_parameters}(f)), while the amplitudes, $A_{\rm H}$, shown in Fig.~\ref{fig:HPspectral_parameters}(d) decrease for low detection energies. 

Figures~\ref{fig:HPspectral}(a) and \ref{fig:HPspectral_parameters}(b) show, that the optical orientation degree measured at the exciton resonance of 1.504~eV does not decrease to zero in a magnetic field of 54~mT. It reaches the level of $P_{\rm oo}(B_{\rm V}=54\, {\rm mT})\approx15\%$, which we assign to the exciton contribution with a width of 160~mT, as reported in Ref.~\onlinecite{kudlacik2024optical}. Indeed, with detuning of the detection energy from the exciton resonance, the $P_{\rm oo}(B_{\rm V}=54\, {\rm mT})$ value gradually decreases down to zero (Fig.~\ref{fig:HPspectral_parameters}(b)). At the detection energy of $E_{\rm det}=1.504$~eV, the PL signal and, therefore, its optical orientation degree, are contributed by the recombination of localized resident carriers and of excitons. At the lower detection energy of $E_{\rm det}=1.498$~eV the exciton contribution is absent. Here, the degree of optical orientation at zero field decreases to about 3\%, and it can be fully suppressed by the Voigt magnetic field of 54~mT, see Figs.~\ref{fig:HPspectral}(b) and \ref{fig:HPspectral_parameters}(b). It is worth to note, that on the basis of the Hanle curves one can distinguish the contributions of excitons and electron-hole pairs to the PL,  for more details see Ref.~\onlinecite{kudlacik2024optical}. 

Even when the nuclear spins are unpolarized and the resulting Overhauser field has zero value, there are the random hyperfine fields of the nuclear spin fluctuations. Depending on their amplitude and the correlation time of the charge carriers the nuclear fluctuations provide carrier spin relaxation, reducing the optical orientation degree. The magnetic field applied in the Faraday geometry eliminates this relaxation process by stabilizing the carrier spins along the field direction. This results in an increase of $P_{\rm oo}$ with growing magnetic field. This effect is known as the polarization recovery (PR) effect~\citep{smirnov2020spin}. 

In the case of free carriers, due to their short correlation time and the weak strength of the nuclear fluctuation fields, this mechanism of spin relaxation is not very efficient~\cite{Dyakonov1973}. For localized carriers the nuclear fluctuation fields can have a magnitude of tens of milliTesla and the correlation time can be long, being dependent on the nature of a localization center~\cite{dyakonov2017}. Depending on the relation between the carrier correlation time and the carrier Larmor precession period, different regimes are established. These regimes are characterized by different ratio between the amplitudes of the Hanle and the polarization recovery curves, as well as their HWHMs ($\Delta_{\rm H}$ and $\Delta_{\rm PR}$)~\cite{kudlacik2024optical, smirnov2020spin, merkulov2002electron, dzhioev2002manipulation}. We thoroughly analyzed their appearance in the lead halide perovskites in Ref.~\onlinecite{kudlacik2024optical}. 

In Fig.~\ref{fig:HPspectral} the PR curves are shown for two detection energies, whether the Hanle curves are measured. The PR curves also have a complex shape and we used three Lorentz contours to fit the data. The fit parameters, $\Delta_{\rm H}$, $\Delta_{\rm PR}$ and the amplitudes ($A_{\rm H}$ and $A_{\rm PR}$) are shown in Fig.~\ref{fig:HPspectral_parameters} as function of energy. One can see, that the HWHM values do not differ in the case of the Hanle and the PR curves (Fig.~\ref{fig:HPspectral_parameters}(e,f)), which evidences that the widths are determined by the nuclear fluctuations. The amplitudes of the Hanle and the PR curves are also close to each other (Fig.~\ref{fig:HPspectral_parameters}(c,d)), which is typical for the interaction of charge carriers with the nuclear spin fluctuations in the limit of intermediate correlation time~\cite{kudlacik2024optical}.

\subsection{Dynamic polarization of nuclear spins}
\label{sec:exp_DNP}

When the sample is illuminated by circularly polarized light, due to the transfer of spin from optically oriented carriers to the lattice nuclei via the hyperfine interaction, dynamic polarization of the nuclei can occur~\cite{meier2012optical,dyakonov2017}. The result of the hyperfine interaction between optically oriented carriers and the nuclei is the creation of the nuclear or Overhauser field of the spin polarized nuclei. For dynamical polarization, the nuclear spins have to be exposed to a magnetic field non-orthogonal to the carrier mean spin, and the circularly-polarized excitation light, used to create the non-equilibrium carrier spin polarization, must be constant in helicity (either $\sigma^+$ or $\sigma^-$).

One of the ways to create and detect the DNP is to use a magnetic field having projection of the excitation light direction, i.e. tilted from the Voigt geometry. The projection of the average spin of the optically polarized carriers on the oblique external field induces the dynamic polarization of the nuclear spins. The resulting Overhauser field $\mathbf{B}_{ N}$, depending on the helicity of the excitation light, is parallel or anti-parallel to the external field. As a result, the spins of the optically oriented carriers are affected by the total field $\mathbf{B}_{\rm tot}=\mathbf{B} \pm \mathbf{B}_{N}$. 

\begin{figure*}
\center{\includegraphics{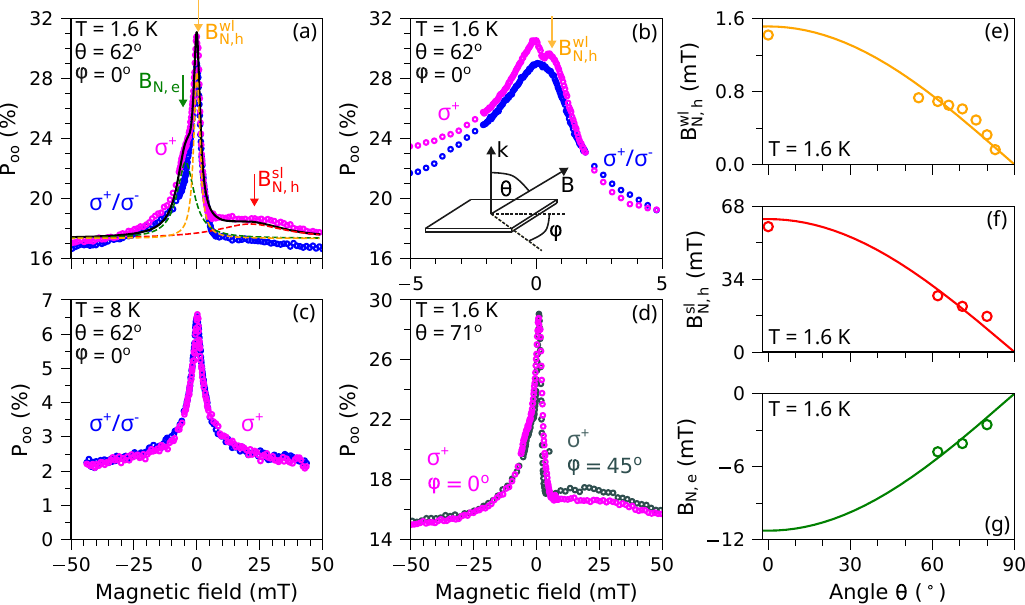}}
\caption{Hanle curves measured in tilted magnetic fields ($\theta = 62^{\circ}$) with $\sigma^+/\sigma^-$ modulated (blue circles) and constant $\sigma^+$ excitation (magenta circles) at (a) $T=1.6$~K and (c) $T=8$~K. $E_{\rm exc}=1.590$~eV,  $P=88$~W/cm\textsuperscript{2}, $E_{\rm det}=1.504$~eV. Arrows indicate the Overhauser field strength for nuclei polarized by the electrons $B_{N,e}$ as well as weakly $B_{N,h}^{\rm wl}$ and strongly $B_{N,h}^{\rm sl}$ localized holes. (b) Expanded view of the low-field region of (a), showing the compensation peak where the Overhauser field from the nuclei polarized by weakly localized holes cancels the external magnetic field. (d) Hanle curves measured for $\theta=71^{\circ}$ and $\sigma^+$ excitation for two in-plane rotation angles of the sample, $\varphi=0^{\circ}$ and $\varphi=45^{\circ}$. (e,f,g) Dependence of the Overhauser field on the tilting angle for weakly and strongly localized holes and for electrons. Circles show the experimental data and lines are fits with a function proportional to the cosine of the angle $\theta$.} 
\label{fig:Hanle_DNP}
\end{figure*}

The Hanle curve measured in the tilted magnetic field at $T=1.6$~K is shown in Fig.~\ref{fig:Hanle_DNP}(a). If the modulated excitation ($\sigma^+/\sigma^-$ at 50~kHz) is used, the nuclei remain unpolarized due to the fast alternation of the mean carrier spin. The resulting Hanle curve has a symmetrical shape with respect to the inversion of the external field, see Fig.~\ref{fig:Hanle_DNP}(a) (blue symbols). 
%\cD{Mladen, but blue symbols give also assymetric Hanle! Confusion! May be can say that small assymetry here is due to not ideal angle alignment...  Kirill, we discuss that with Mladen. W edo not know the reason for assymetry.  Our suggestion to not to dicuss that and claim that it is symmetric. But may be you have idea what we can say here....} \cK{We discussed this with Mladen some time ago, and also did not find a good solution} 
Otherwise, for excitation with fixed circularly polarized light, the nuclei polarization is possible, which is evident by the asymmetric Hanle curve due to the presence of the Overhauser field, see magenta symbols  in Fig.~\ref{fig:Hanle_DNP}(a). The three shoulders that appear at $B_{N,h}^{\rm sl}\approx24$~mT, $B_{N,e}\approx-5$~mT, and $B_{N,h}^{\rm wl}\approx0.7$~mT in Fig.~\ref{fig:Hanle_DNP}(a) are a consequence of the compensation of the external field by the Overhauser field. To enhance visibility, a zoom of the shoulder at $B_{N,h}^{\rm wl} \approx 0.7$~mT is provided in Fig.~4(b). %\cD{Colleagues, I am missing, where we give assignement of components to electrons and holes. We may refer our previous paper on that, but better to repeat main arguments... Let us check that.}

The spin polarizations of electron $S_{e}$ and hole $S_{ h}$ are aligned with the $k$-vector of light, following the selection rules for $\sigma^+$ excitation. The resulting nuclear spin polarization $\langle I_{N} \rangle$ will also align along the same direction. The polarity of Overhauser fields created by polarized spins of the same nuclear species for the two types of charge carriers is determined by the signs of carrier $g$-factors. Since $g_{ e} = +3.57$ for the electrons, and $g_{h} = -1.21$ for the holes~\cite{kirstein2022lead}, the corresponding nuclear fields for electrons and holes have opposite signs.
%\cD{We do not have such letter in theory part. Kirill, please check. Can we establish relation with theory here? Or we will do that later in Discussion?} and on the signs of the electron and hole $g$-factors, $B_{N,e(h)}=A_{ e(h)} \langle I_{N} \rangle / \mu_{B} g_{e(h)}$. Since $g_{ e} = +3.57$ for the electrons, and $g_{h} = -1.21$ for the holes~\cite{kirstein2022lead}, the corresponding nuclear fields for electrons and holes have opposite signs. 

In order to verify that the observed compensation shoulders are indeed related to the presence of the Overhauser fields, we measured Hanle curves at the different tilting angles of the external magnetic field. It is clearly seen from Figs.~\ref{fig:Hanle_DNP}(e), ~\ref{fig:Hanle_DNP}(f) and ~\ref{fig:Hanle_DNP}(g), that the magnitudes of the Overhauser fields resulting from the interaction with holes and electrons decrease as the angle becomes larger and approaches the Voigt geometry. The specific of lead halide perovskites is that the electrons have a much weaker hyperfine interaction with the nuclei than the holes, which results in a smaller Overhauser field. Accordingly, the Overhauser field of $B_{N,h}^{\rm sl} = 26$~mT at $\theta=62^\circ$ is attributed to the interaction between the lattice nuclei and the hole, whereas the field of $B_{N,e} = -5$~mT at the same angle is associated with the interactions between the nuclei and the electron. As a result, we attribute the contributions to the Hanle and PR curves with half-widths of 20~mT and 5~mT to localized holes and electrons, respectively, as demonstrated in our previous work~\cite{kudlacik2024optical}. 

The compensating field $B_{N,h}^{\rm wl} = 0.7$~mT shown in Fig.~\ref{fig:Hanle_DNP}(b) has the same sign as $B_{N,h}^{\rm sl}=26$~mT, i.e. the Overhauser field of the localized holes, although its magnitude is considerably smaller. This difference can be explained by the coexistence of weakly and strongly localized holes in the studied crystal. The strongly localized holes transfer their spin to the nuclei more efficiently, resulting in a larger Overhauser field. Moreover, they couple more efficiently to the nuclear spin fluctuations, which leads to broader Hanle and PR curves. In contrast, the weakly localized holes experience smaller fluctuations and, therefore, have narrower Hanle and PR curves. Consequently, we attribute the narrow peaks in the Hanle and PR curves with half-widths on the order of 0.5~mT to weakly localized holes with the resulting Overhauser field of $B_{N,h}^{\rm wl} = 0.7$~mT ($\theta=62^\circ$).

The DNP effect is strongly temperature dependent. For the Hanle curves measured at $T=8$~K one cannot identify a difference between the excitation with light of constant ($\sigma^+$) and modulated ($\sigma^+/\sigma^-$) polarization, see Fig.~\ref{fig:Hanle_DNP}(c). A possible explanation for this behavior is that at higher temperatures the carrier delocalization occurs, which results in their rather weak interaction with the nuclei. This reduced coupling diminishes the efficiency of dynamic nuclear polarization, leading to weaker nuclear spin effects on the electron and hole spins.

\subsection{Temperature dependence of the Hanle and polarization recovery curves}
\label{sec:exp_temperature}

%\cD{Colleagues, I have reworded  this section. Ideas are kept, but I tryed to express them more clear and easier. Please check.}

\begin{figure}[t!]
\center{\includegraphics{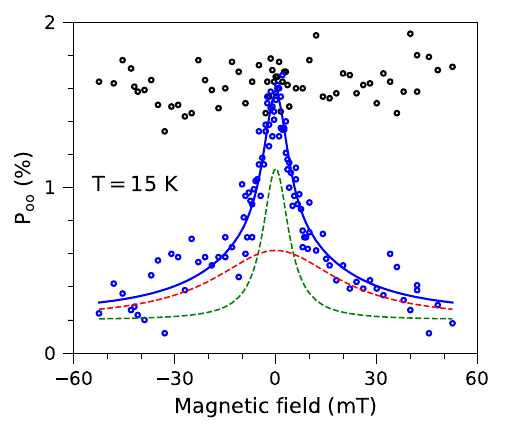}}
\caption{Polarization recovery (black circles) and Hanle (blue circles) curves measured at $T=15$~K with modulated $\sigma^+/\sigma^-$ excitation. $E_{\rm exc}=1.590$ eV, $P=18$ W/cm\textsuperscript{2}, $E_{\rm det}=1.504$ eV. The solid blue line is a fit with the sum of two Lorentz functions, whose individual components are shown by the dashed green line for electrons ($\Delta_{N,e}=5$~mT) and the dashed red line for strongly localized holes ($\Delta_{N,h}=23$~mT). 
}
\label{fig:Hanle+PRC}
\end{figure}

The hyperfine interaction of localized carriers with the nuclear spin fluctuations depends sensitively on the lattice temperature, whose increase induces carrier delocalization and shortens of their correlation time. One can see, that by comparing the Hanle and PR curves at $T=1.6$~K in Fig.~\ref{fig:HPspectral}(a) with those at $T=15$~K shown in Fig.~\ref{fig:Hanle+PRC}. At zero magnetic field the degree of optical orientation decreases from 25\% to 3\% evidencing more efficient carrier spin relaxation at elevated temperatures. The Hanle curve at $T=15$~K can be fitted with two Lorentzian functions with $\Delta_{N,e}=5$~mT and $\Delta_{N,h}=23$~mT. The narrowest component, which at $T=1.6$~K has the width of $\Delta_{N,h}^{\rm wl}=0.3$~mT is absent at $T=15$~K. The PR curve becomes flat and shows no dependence on the Faraday magnetic field in this range. 

\begin{figure*}[t!]
\center{\includegraphics{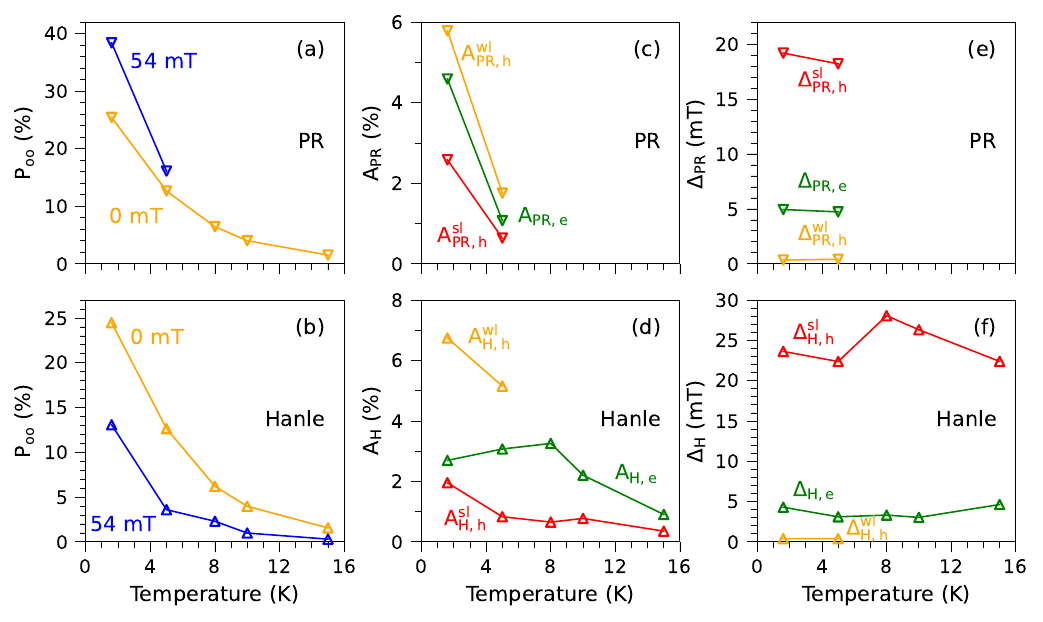}}
\caption{Experimental parameters extracted from the Hanle and polarization recovery curves recorded at various temperatures. $\sigma^+/\sigma^-$ excitation mosulated at 50 kHz with $E_{\rm exc}=1.590$~eV, $P=18$~W/cm\textsuperscript{2}, $E_{\rm det}=1.504$~eV. (a,b) Temperature dependence of the optical orientation degree at zero field and at 54 mT. Comparison of the (c,d) amplitudes of the three Lorentz contours and their (e,f) HWHMs evaluated from the Hanle and PR curves at various temperatures. In the plots, the amplitudes and half-widths of strongly localized holes, electrons, and weakly localized holes are plotted by red, green, and orange symbols, respectively.}
\label{fig:Hanle+PRC+Temp+Parameters}
\end{figure*}

In order to study these changes in detail, we measured the Hanle and PR curves in the temperature range of $1.6-15$~K. The obtained parameters are presented in Fig.~\ref{fig:Hanle+PRC+Temp+Parameters}. Several temperature ranges can be distinguished.

For temperatures from 1.6 to 6~K, the Hanle and PR curves contain three components so that they can be fitted with the sum of three Lorentzians. Their HWHM values differ, and we assign the components from the largest to the smallest width to strongly localized holes,  electrons, and weakly localized holes. The HWHMs for the Hanle ($\Delta_{\rm H}$) and PR ($\Delta_{\rm PR}$) curves are almost identical, as shown in Figs.~\ref{fig:Hanle+PRC+Temp+Parameters}(e) and \ref{fig:Hanle+PRC+Temp+Parameters}(f). This behavior is typical for carriers interacting with the nuclear spin fluctuations in the regime of long correlation times~\cite{dzhioev2002manipulation, merkulov2002electron}. 

For temperatures above 6~K, the PR amplitude, $A_{\rm PR}$, becomes zero, see Fig.~\ref{fig:Hanle+PRC+Temp+Parameters}(c). This can be explained by delocalization of the electrons and holes. As mentioned above, the Overhauser fields of the nuclear spin fluctuations can lead to carrier spin relaxation. However, for free carriers this relaxation mechanism is inefficient, since these fields are weak and the carrier correlation time with nuclear  spin fluctuations is short. Additionally, the DNP created by optically oriented carriers is more efficient when they are localized. By comparing the Hanle curves measured in a tilted magnetic field and for constant $\sigma^+$ excitation at $T=1.6$~K and 8~K, one can see that at 1.6~K (Fig.~\ref{fig:Hanle_DNP}(a)), when the carriers are localized, the Hanle curve has an asymmetric shape due to the presence of the Overhauser field, while at $8$~K the Hanle curve is symmetric (Fig.~\ref{fig:Hanle_DNP}(c)) evidencing the delocalizazion of carriers.  

At temperatures higher than 8~K,  the Hanle curves are well described by the sum of two Lorentzians, with $\Delta_{N,e}\approx5$~mT and $\Delta_{N,h}\approx20$~mT, see Figs.~\ref{fig:Hanle+PRC} and \ref{fig:Hanle+PRC+Temp+Parameters}(f). The narrow component with $\Delta_{N,h}^{\rm wl}\approx0.3$~mT, which has large amplitude at $T=1.6$~K  (Fig.~\ref{fig:HPspectral}(a)) is absent for $T>8$~K, which can be explained by the delocalization of the weakly localized holes in this temperature range. 
%By comparison with the Hanle curve measured at 1.6~K (Fig.~\ref{fig:HPspectral}(a)), it can be seen that the narrow contour with a HWHM of 0.4~mT is absent when $T\geqslant8$~K. This fact indicates that the weakly localized holes are delocalized at these temperatures. 

It is important to note, that at $T<8$~K pronounced Hanle and PR curves are observed, which have the same HWHM. From that we conclude that in this temperature range the electron and hole spin relaxation is mainly provided by their hyperfine interaction with the nuclear spin fluctuations. Above 8~K the PR effect vanishes, namely the PR amplitude decreases while the FWHM is maintained, see Figs.~\ref{fig:Hanle+PRC+Temp+Parameters}(c,e). This means that for delocalized carriers the hyperfine interaction is not the leading spin relaxation mechanism anymore. The Hanle curve decreases in amplitude, but still has a finite value, while the HWHM of the Hanle components do not change in the range of $1.6-15$~K, see Figs.~\ref{fig:Hanle+PRC+Temp+Parameters}(d,f). This evidences that at $T>8$~K the Hanle HWHM is 
%Below 8~K, the interaction of localized carriers with nuclear spin fluctuations was the main mechanism of spin relaxation of holes and electrons with characteristic $\Delta_{\rm H}$ and $\Delta_{\rm PR}$ for Hanle and PR curves. Since the polarization recovery effect is missing at higher temperatures, the $\Delta_{\rm H}$ of the Hanle curve will no longer be linked to the magnitude of the nuclear fluctuation field, but it will be 
determined by the value of the carrier spin lifetime $T_{s}$. In its classical form, the Hanle effect is described by~\cite{meier2012optical}:
\begin{equation}
S_{z}(B)=\frac{S_{z(0)}}{1+(B / \Delta_{{\rm H},e(h)})^2}.
\label{eq:Hanle}
\end{equation}
Here, $S_{z(0)}$ is the projection of the spin on the direction of the initial spin polarization $S_{ 0}$ ($z$-axis) and $B$ is the external magnetic field applied in the Voigt geometry. One can see from Eq.~\eqref{eq:Hanle} that the depolarization of the carrier spin in a transverse magnetic field is well described by a Lorentz function. The characteristic field $\Delta_{{\rm H},e(h)}$ describes the HWHM of the electron (hole) contribution to the Hanle curve and is inversely proportional to their spin lifetime $T_{s,e(h)}$:
\begin{equation}
\Delta_{{\rm H},e(h)}=\frac{\hbar}{\mu_{B} g_{e(h)} T_{s,e(h)}},
\label{eq:Hanle2}
\end{equation}
where $\hbar$ is the reduced Planck constant, $\mu_{B}$ is the Bohr magneton, and $g_{e(h)}$ is the electron(hole) $g$-factor. The spin lifetime itself is limited by the recombination $\tau_{e(h)}$ and spin relaxation $\tau_{s,e(h)}$ times: $1/T_{s,e(h)}=1/\tau_{e(h)}+1/\tau_{s,e(h)}$. The Hanle curve half-widths measured at $T=15$~K are $\Delta_{{\rm H},h}=23$~mT and 
$\Delta_{{\rm H},e}=5$~mT. Using Eq.~\eqref{eq:Hanle2} we evaluate the corresponding spin lifetimes of $T_{s,h}=0.41$~ns for the holes and $T_{s,e}=0.64$~ns for the electrons.

\subsection{Dynamics of nuclear spin-lattice relaxation}
\label{sec:exp_T1}

%\cD{Colleagues, I implement rewording in this section. with the goal to simplify it for readers. Please check.}

The spin-lattice relaxation of  nuclear spins due to their interaction with localized carriers was studied using a three-step experimental protocol that includes optical pumping and measurements of the difference in the nuclear spin polarization after a dark interval of variable length. A similar procedure was used in Ref.~\cite{kotur2014nuclear}. In the first stage, the nuclei are polarized through their interaction with optically oriented carriers in an oblique external magnetic field. The Overhauser field of the polarized nuclei changes the optical orientation degree of the carriers, which can be used for detecting the nuclear spin dynamics.

For the first step, we used pumping pump duration of $t_{\rm pump}=90$~s, based on the observation that the optical orientation reaches a plateau during this time scale, indicating that the influence of the polarized nuclei has reached its maximum effect for the selected pumping conditions. For this, we used fixed circularly polarized light ($\sigma^+$) with the power density of $P=88$~W/cm\textsuperscript{2}. An external magnetic field of $B_{\rm ext}=-3.5$~mT was applied at the angle of $\theta=62^\circ$, where a shoulder appears in the Hanle curve (Fig.~\ref{fig:Hanle_DNP}(a)), corresponding to the compensation of the external field by the Overhauser field generated by the spin-polarized nuclei through their interaction with the electrons.

The second step begins with the pump being blocked by a shutter for an arbitrary time $t_{\rm dark}$, which was varied from 1 to 40~s. During this step, the nuclear spin polarization decays as $\exp{(-t_{\rm dark}/T_{\rm 1,dark})}$ due to spin-lattice relaxation process with the nuclear spin-lattice relaxation time $T_{\rm 1,dark}$. 

\begin{figure*}
\center{\includegraphics{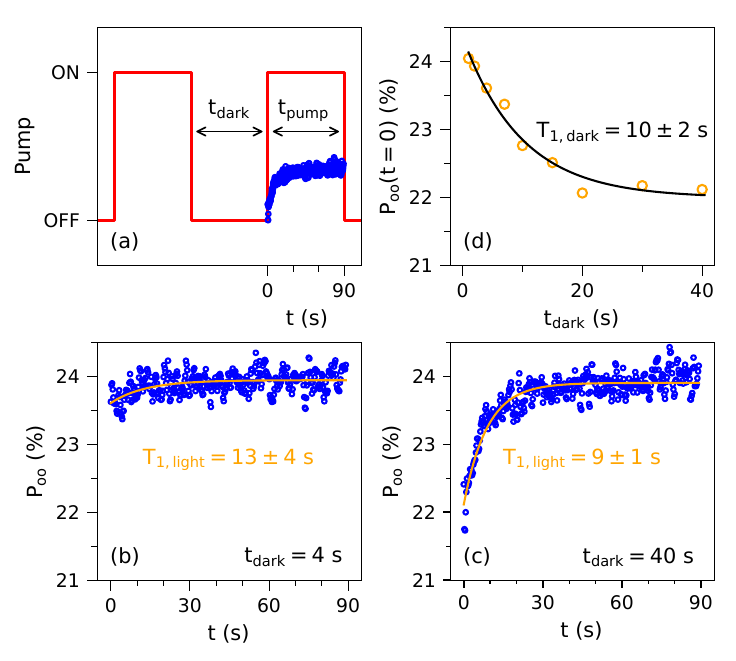}}
\caption{(a) Timeline of the experimental protocol used to measure the nuclear spin-lattice relaxation time at $T=1.6$~K. The sample was pumped for $t_{\rm pump}=90$~s with an excitation power of $P = 88$~W/cm\textsuperscript{2} in an external magnetic field of $B_{\rm ext}=-3.5$~mT, tilted at an angle of $\theta=62^\circ$. Subsequently, it was kept in the dark for various delay times between $t_{\rm dark}=1$~s and $40$~s. (b,c) Build-up of the nuclear spin polarization measured via the $P_{\rm oo}(t)$ dynamics after $t_{\rm dark}=4$~s and $40$~s. 
%The PL circular polarization degree was recorded as a function of time $t$, where 
$t=0$ corresponds to the start of the measurement immediately after opening the laser beam up to $t=t_{\rm pump}=90$~s, with data points redorded every 0.2~s. Fitting these data with an exponential function (orange lines) gives the nuclear spin relaxation time under illumination, $T_{\rm 1,light}$. The initial value of each fit, $P_{\rm oo}(t=0)$, reflects the remaining nuclear spin polarization after the preceding dark period, and its dependence on $t_{\rm dark}$ given in panel (d) shows how the nuclear polarization decays in darkness. The solid black line is an exponential fit to these points, yielding the nuclear spin-lattice relaxation time $T_{\rm 1,dark}$.} 
\label{fig:DE}
\end{figure*}

In the third step, the degree of optical orientation is recorded as function of time after opening the pump again. The measurement starts at $t=0$ and continues up to $t=t_{\rm pump}=90$~s, with an accumulation time of 0.2~s per point. Similar to the initial first step, the duration of 90~s is chosen to ensure complete restoration of the nuclear spin polarization under the selected pumping conditions. Figures~\ref{fig:DE}(b) and \ref{fig:DE}(c) show examples of how the degree of optical orientation evolves after dark periods of 4 and 40~s. By fitting these dependences with an exponential function, we evaluate the nuclear spin-lattice relaxation time under illumination, $T_{\rm 1,light}$, which is needed for the nuclear spin polarization to be restored. Also, the initial value of the optical orientation, $P_{\rm oo}(t=0)$, recorded immediately after unblocking the laser, reflects the strength of the remaining nuclear field and depends on the duration of $t_{\rm dark}$. As a result, we obtain the dependence of the Overhauser field on the duration of the second step, which can be well described by an exponential function with the spin-lattice relaxation time $T_{\rm 1,dark}$ as fit parameter, see Figure~\ref{fig:DE}(d). For improved accuracy, the sequence of the $t_{\rm dark}$ and $t_{\rm pump}$ intervals is repeated 10 times, and the optical orientation degree measured after each sequence is averaged (Figure~\ref{fig:DE}(a)). One can see, that the times $T_{\rm 1,light}=13$~s after $t_{\rm dark}=4$~s and $T_{\rm 1,light}=9$~s after $t_{\rm dark}=40$~s, required to restore the nuclear spin polarization under illumination, match the nuclear spin-lattice relaxation time in the dark, $T_{\rm 1,dark}=10$~s. The measured times are also reasonably close to the reference value of $T_{\rm 1,light}=6$~s, obtained under optical pumping by means of time-resolved Kerr rotation measurements on the same sample~\citep{kirstein2022lead}.

%\begin{table}
%\centering
%\caption{Nuclear spin relaxation times in the dark and under illumination. $T=1.6$ K. \cD{remove data at 4.2 mT}    \cD{Mladen, I think that for one field we do not need Table. Just remove it and give reference value in text. As well as other two times.} }
%\begin{tabular}{|c|c|c|c|}
%\hline
% & $B_{\rm dark} = 3.5$~mT & $B_{\rm dark}=4.2$~mT & \shortstack{Reference \\ value~\cite{kirstein2022lead}} \\ \hline
% & 3.5 & 4.2 & value~\cite{kirstein2022lead} \\ \hline

%$T_{\rm 1,light}$~(s) & 9 & 8 & 6 \\ \hline
%$T_{\rm 1,dark}$~(s) & 10 & 11 & - \\ \hline
%\end{tabular}
%\label{tab:samples}
%\end{table}

%%%%%%%%%%%%%%%%%%%%%%%%%%%%%%%%%%%%%%%%%%%%%%%%%%%%%%%%%%%%%%%%%%%%%%%%%%
\section{Discussion}
\label{sec:discussion}

\subsection{Overhauser fields for electrons and holes}
\label{sec:discussion_Overhauser}

Measuring the Hanle effect in oblique external magnetic field allows one to determine the Overhauser fields acting on the spins of charge carriers. In the FA$_{0.9}$Cs$_{0.1}$PbI$_{2.8}$Br$_{0.2}$ perovskite semiconductor studied here, the distinction between the Overhauser fields acting on the electrons and the holes is simplified by the different signs of their $g$-factors. As a result, compensation of the Overhauser fields for the electrons and the holes is reached for different polarities of the external magnetic field, so that their Overhauser fields can be measured independently. 

The contributions of the lead and halogen nuclei to these Overhauser fields can be distinguished by their different angular dependences. Because of the anisotropic hyperfine coupling of the halogen nuclei with both the electrons and the holes, the dependence of the Overhauser fields on the polar and azimuthal angles of the external field in principle might be complex (see Section \ref{subsec:DNP theory}), following the cubic symmetry.  However, as one can see from Figs.~\ref{fig:Hanle_DNP}(e,f,g), the Overhauser field as function of the polar angle $\theta$ follows a simple cosine dependence, typical for an isotropic hyperfine coupling. In addition, there is virtually no difference between the Hanle curves measured for the azimuthal angle set to $\varphi=0$ (supposedly, corresponding to the external field oriented along the [110] cubic axis) and to $\varphi=\pi/4$ (Fig.~\ref{fig:Hanle_DNP}(d)).  This controversy is more illustrative in the case of the conduction band, because the electrons are more strongly coupled with the halogen nuclei than the holes.  

\begin{figure}
 \includegraphics{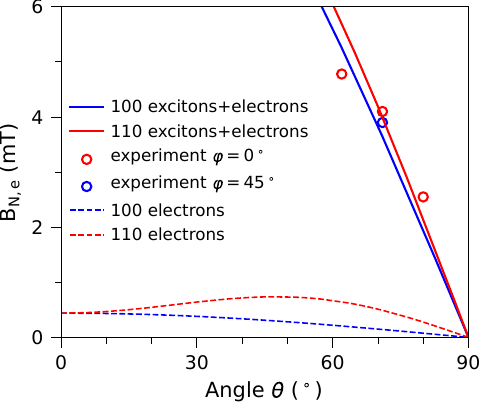}
%\center{\includegraphics{Figure_BN_electrons.pdf}}
\caption{Dependences of the Overhauser field for conduction-band electrons on the azimuthal angle $\theta$ calculated  for two orientations of the in-plane component of the external magnetic field, along [110] (red) and [100] (blue). Dashed lines: DNP of lead and iodine by electrons with the spin polarization of 9\%. Solid lines: DNP of lead up to 63\% by excitons and electrons, while iodine is polarized by electrons only. Circles: Overhauser fields for electrons from experiment, see Fig.~\ref{fig:Hanle_DNP}(g). }  
\label{fig:BN electrons}
\end{figure}

The theoretical dependences of the Overhauser field on $\theta$, calculated using Eq.~\eqref{eq:eq31}, are plotted in Fig.~\ref{fig:BN electrons} by the dashed lines for the in-plane component of the external field oriented along the [110] and [100] axes. In these calculations we used the hyperfine constant of iodine $A^s_I|C_{cs}|^2\approx17\,\mu$eV, experimentally determined in Ref. \cite{Meliakov2025}, as well as the hyperfine constant of lead $A^{cbb}_{Pb}\approx40\,\mu$eV~\cite{Meliakov2025}, estimated on basis of earlier measurements for a p-type Bloch amplitude (valence band in PbTe~\cite{hewes1973nuclear}). The electron spin polarization of 9\% was used, defined as the sum of the amplitudes of the Hanle and PR curves (Figs.~\ref{fig:HPspectral_parameters}(c,d)). The theoretical curves, shown by the dashed lines, clearly disagree with the experimental data given by the circles. First, the experiment does not exhibit a considerable difference between the Overhauser fields measured at different $\varphi$. Second, the theoretical Overhauser fields are several times smaller than the experimentally measured values for the corresponding angles of the external field.

 \begin{figure}
  \includegraphics{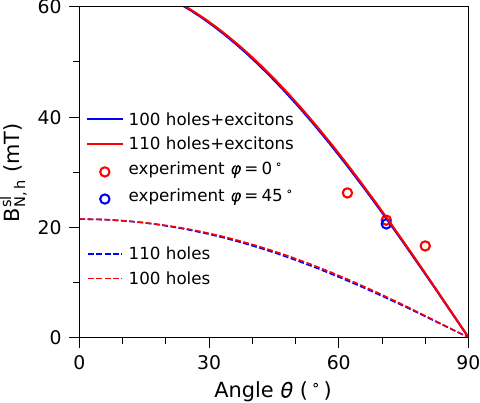}
%\center{\includegraphics{Figure_BN_holes.pdf}}
\caption{Dependences of the Overhauser field for valence-band holes on the azimuthal angle  calculated  for two orientations of the in-plane component of the external field, along [110] (red) and [100] (blue). Dashed lines: DNP of lead and iodine by holes with the spin polarization of 5\%. Solid lines: DNP of lead up to 15\% by excitons and holes, while iodine is polarized by holes only. Circles: Overhauser fields for strongly localized holes from experiment, see Fig.~\ref{fig:Hanle_DNP}(f).} \label{fig:BN holes}
\end{figure}
 
Turning to the Overhauser field for the holes, one notices that the anisotropy of its dependence on the in-plane angle $\varphi$ is small even in theory, because of the weak hyperfine interaction of the halogens with the holes. This is clearly seen in Fig.~\ref{fig:BN holes}, where the dependences calculated for the [110] and [100] crystal orientations closely follow each other. However, the absolute value of the Overhauser field, calculated from the hole spin polarization of 5\%, measured as the sum of the amplitudes of the Hanle and PR curves (Figs.~\ref{fig:HPspectral_parameters}(c,d)), is about 3 times smaller than the one measured experimentally, see Fig.~\ref{fig:Hanle_DNP}(f) and the symbols in Fig.~\ref{fig:BN holes}. Note that the hyperfine constant for the hole interaction  with the lead spins, $A^{vbt}_{Pb}\approx280\,\mu$eV, was experimentally measured for CsPbBr$_3$ nanocrystals \cite{Meliakov2024}. This value is close to the atomic hyperfine constant (see Table \ref{tab:hyperfineparam}) and we suggest that it can be safely used for the studied FA$_{0.9}$Cs$_{0.1}$PbI$_{2.8}$Br$_{0.2}$ crystals. 

In order to achieve better agreement of the theory with the experimental data, while being restricted by the previously determined hyperfine constants, we have to assume that the nuclear spins are polarized more strongly than can be provided by their interaction with localized charge carriers. One should note that a proportional increase of the spin polarization of both the lead and the iodine nuclei would only increase the magnitude of the angular dependences of the Overhauser fields, shown in Figs.~\ref{fig:BN electrons} and \ref{fig:BN holes} by the dashed lines, while leaving the shape unchanged. This strongly contradicts the experiment (at least for the electrons). Therefore, one should look for a source of additional DNP, affecting mainly the spins of the lead nuclei.

 %Having considered the collection of cw and time-resolved spectra of polarized photoluminescence, accumulated in the literature, \cD{give references} one can suggest that this excessive polarization  of lead nuclei could result from their interaction with spin-oriented excitons. 
 
We suggest that this excessive polarization of the lead nuclei may result from their interaction with spin-oriented excitons.  Indeed, the excitons in bulk perovskite crystals exhibit a very high (over 80\%) circular polarization in their emission, induced by optical orientation through circularly polarized excitation~\cite{kopteva2023giant}. However, the exciton lifetime is short (55~ps), as is their spin relaxation time (300~ps)~\cite{kopteva2023giant}. These times are much shorter than the precession periods of the electron and hole spins in the effective magnetic fields of the nuclear spin fluctuations on the order of 10~ns~\cite{kudlacik2024optical}. Therefore, the DNP induced by the excitons occurs in the regime of short spin correlation time. In this regime, the rate of spin transfer to the nuclear spin system is proportional to the squared hyperfine constant~\cite{OOChapter2, Dyakonov1973} and for this reason it is much larger for lead than for iodine. Specifically, for the DNP via electron spin-flips the difference is about a factor of 50, while for the DNP via holes it reaches about four orders of magnitude. This would result in a strong spin polarization of the lead nuclei, while for the spins of iodine pumping by excitons might not be strong enough to overcome spin leakage. The mean spin of the iodine nuclei would then be determined by the smaller mean spin of the localized charge carriers, which have long spin correlation times \cite{kudlacik2024optical} and can effectively polarize the spins of all nuclear species. 

The enhanced DNP of lead results in a larger coefficient $a_{44}$ in Eq.~\eqref{eq:eq31}. Using $a_{44}$ as a fitting parameter, a reasonably good agreement of the theoretical angular dependences of the Overhauser field with the experiment can be achieved. The solid lines in Figs.~\ref{fig:BN electrons} and~\ref{fig:BN holes} are plotted for spin polarizations of the lead nuclei equal to 63\% and 15\%, respectively.

The different excess polarization of lead, deduced from the Overhauser fields acting on the localized electrons and holes can be explained by the following mechanism of DNP mediated by charged excitons (trions). A spin-oriented exciton relaxes in energy without losing its spin orientation and is captured into a localization site where one of the long-living charge carriers is already localized. As a result, a localized trion is formed. In case of a resident electron, the two electrons in teh trion form a singlet state, while the hole remains an uncompensated spin that can efficiently polarize the spins of the lead nuclei because of their strong hyperfine coupling with holes. Similarly, in the case of a resident hole a positively-charged trion is formed, and the DNP of the lead nuclei is mediated by the electron from the captured exciton. As the hyperfine coupling of the lead nuclei with the electrons is weaker than with the holes, the DNP in this case is less efficient. This can explain the weaker spin polarization of lead deduced from the Overhauser field for resident holes (15\%) as compared to that for resident electrons (63\%).

\subsection{DNP near zero magnetic field}
\label{sec:discussion_DNP_zero_field}

\begin{figure*}
\center{\includegraphics{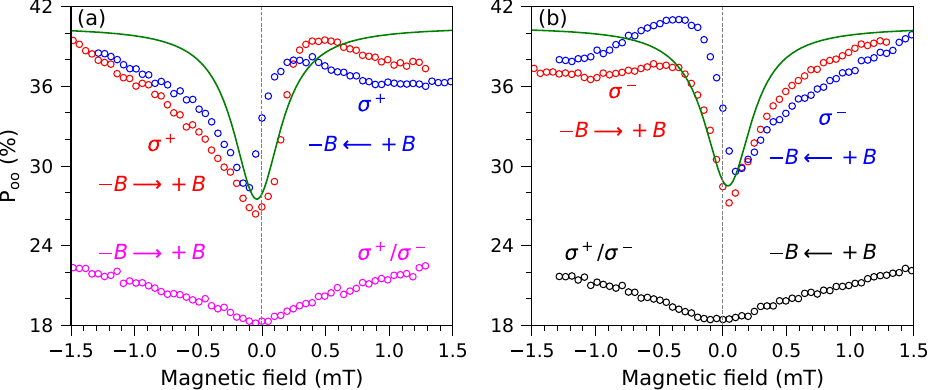}}
\caption{Zero-field peak shift in the PR curves, measured in Faraday geometry at $T = 1.6$~K with $E_{exc}=1.590$~eV and $E_{det}=1.504$~eV, for different signs of the excitation polarization: (a) $\sigma^+$ and (b) $\sigma^-$. The circles give experimental data for two directions of the magnetic field sweep: red – from negative to positive and blue – from positive to negative. Solid lines are Lorentzian fits with a HWHM 
%\cD{is that HWHM?} 
of 0.23~mT, centered at $\pm0.057$~mT. For comparison, the experimental data for modulated $\sigma^+ / \sigma^-$ excitation are also shown for the two sweep directions: magenta - from negative to positive and black – from positive to negative. The dashed vertical lines indicate the zero magnetic field position.}
\label{fig:ZeroFieldPeak}
\end{figure*}

Since the Overhauser fields for both electrons and holes are dominated by the polarized spins of the lead nuclei, one should expect them to exhibit the behavior in weak external fields that is typical for nuclear spins without quadrupole splitting~\cite{OOChapter5}. Such spins need a magnetic field to support their polarization gained via DNP. Their mean spin decreases to zero at zero external field, and increases quadratically with the external field on the scale of the local magnetic fields of the dipole-dipole interactions. This property manifests itself as a narrow dip on the PR curve or a narrow peak on the oblique-field Hanle curve~\cite{OOChapter5}. These features are indeed observed in our experiments in Figs.~\ref{fig:Hanle_DNP} and \ref{fig:ZeroFieldPeak}. A closer look at the PRC reveals that for unmodulated excitation light, the minimum of the curve near zero magnetic field is shifted by $0.05–0.1$~mT compared to the case with laser helicity modulation (data points in Fig.~\ref{fig:ZeroFieldPeak}). This shift reverses its sign when the polarization of the excitation light is inverted. The PR curves recorded for different scan directions are slightly shifted towards opposite signs of the magnetic field because of the finite response time of the nuclear spin system (Fig.~\ref{fig:DE}), but both curves demonstrate the same signs of the $\sigma$-dependent shifts. This suggests the presence of some internal field acting on the nuclear spins, which arises from optical spin orientation and compensates the external field at the position of the PL polarization dip.  

At first glance, this shift might be taken to be the well-known effect of the Knight field on the polarized charge carriers \cite{OOChapter5}. However, a more accurate consideration rules out this assignment. Indeed, exposing the sample to $\sigma^+$ polarized light leads to a positive sign of the $z$-component of the carrier spin. Following Ref.~\onlinecite{OOChapter5} one can see, that the direction of the Knight field is opposite to that of the average spin. Then, the shift of the minimum under $\sigma^+$ excitation arises from the compensation of the negative Knight field, and should occur at a positive external field strength.  In fact, in our sample the minimum shifts in the opposite direction, see Fig.~\ref{fig:ZeroFieldPeak}(a). 

We suppose that the observed shift of the zero dip of the PR curve is due to the weak magnetic field generated by the polarized spins of the halogen nuclei at the lead nuclei. According to Eqs.~\eqref{eq:eq26} and \eqref{eq:eq27}, there is a field-independent polarization of the quadrupole-split spins of the halogen nuclei that is governed solely by the average carrier spin. As follows from Eqs.~\eqref{eq:eq6}-\eqref{eq:eq8}, the polarized halogen nuclei produce a small magnetic field acting on both the lead and halogen nuclei. A comparison of the coefficients $L_1 – L_3$ shows that the strongest field will be generated at the lead nuclei. When this field is compensated, the dynamic nuclear polarization of the lead nuclei is suppressed, resulting in a minimum on the PR curve. 

Using the values of the electron and hole polarization obtained in the experiment, one can estimate the magnitude of the fields and, consequently, the positions of the minimum on the PR curve. Taking values of 10\% for the electron and 5\% for the hole spin polarization, we obtain the values of the field created by the iodine spins acting on the lead nuclei as 0.057 and 0.029~mT, respectively. The half-width of the dip is determined by the fluctuations in the dipole-dipole nuclear fields at the lead nuclei (Eq.~\eqref{eq:eq5}), yielding a value of 0.23~mT for $\text{FAPbI}_3$. Since the electron spin polarization gives the larger contribution to the PL polarization, we plot it in Fig.~\ref{fig:ZeroFieldPeak} for comparison with the experimental data. Theoretical curves, modeled as Lorentzians with 0.23 mT half-width and centered at $-0.057$ and $+0.057$~mT, are plotted as solid lines in Fig.~\ref{fig:ZeroFieldPeak}(a) for $\sigma^+$ excitation and in Fig.~\ref{fig:ZeroFieldPeak}(b) for $\sigma^-$ excitation. One can see, a reasonably good agreement between the theory and the experimental data.

Yet, another manifestation of the DNP of the halogens is the fact that the optical orientation measured under modulated excitation is notably smaller than the minimum polarization under excitation with constant $\sigma^+$ or $\sigma^-$  polarization (Fig.~\ref{fig:ZeroFieldPeak}). 
%\cD{Kirill, do we refer here results in Fig.~\ref{fig:ZeroFieldPeak}? I would them add it here.} 
This evidences the existence of a contribution of the quadrupole-split iodine spins to the Overhauser field, which is independent of the magnetic field acting on the nuclear spins (Eqs.~\eqref{eq:eq28} and \eqref{eq:eq29}) and can reach several milliTesla for both electrons and holes. It is worth to note that the anisotropy of the Hanle curves, expected due to the DNP of the halogens, is not pronounced in our experiments because of the excess polarization of the Pb spins by excitons (see the previous subsection). Under these circumstances, the observation of the zero dip shift of the PR curves is an important indication of the DNP of the halogen spins and their interaction with the charge carriers.

\section{Conclusions}
\label{sec:conclusions}

The comparison of the general theory, that we have developed for cubic lead halide perovskites, with the experiment performed on a FA$_{0.9}$Cs$_{0.1}$PbI$_{2.8}$Br$_{0.2}$ crystal allows us to outline the overall pattern of the DNP in the perovskite semiconductors. It includes the dynamic spin polarizaton of both the lead and halogen nuclei by localized charge carriers, the additional polarization of the lead spins by excitons, and the magnetic field created by the quadrupole-split halogen spins at the lead nuclei spins. Specific predictions of the theory, like the anisotropic Hanle effect due to the Overhauser fields of the halogens, did not occur in a pronounced way in the present experiments, but could manifest themselves more strongly in other perovskite crystals. The experiment reveals a bunch of unexpected effects, like the narrow peak due to weakly localized holes in the Hanle and PR curves, which deserve further investigation. The quenching of the DNP when elevating othe temperature to just 8~K, as well as the 10-second time scale of the spin-lattice relaxation both under optical pumping and in the dark, also require a theoretical explanation. The generality of the experimental findings needs to be tested and confirmed by studies of other lead halide perovskite crystals, e.g., with Br and Cl halogens.

\section*{Acknowledgments} 
The authors are thankful to V. L. Korenev, M. M. Glazov, I. A. Akimov, and E. Kirstein for fruitful discussions. M.K. and M.B. acknowledge support by the BMBF project QR.N (Contract No.16KIS2201). D.R.Y. acknowledges support by the Deutsche Forschungsgemeinschaft via the SPP2196 Priority Program (Project YA 65/28-1, No. 527080192). N.E.K. acknowledges the support of the Deutsche Forschungsgemeinschaft (Project KO 7298/1-1, No. 552699366). K.V.K. acknowledges support by the Saint Petersburg State University (Grant No. 125022803069-4). The samples for this study were provided by O. Hordiichuk, D. N. Dirin, and M. V. Kovalenko from ETH Z\"urich.

%%%%%%%%%%%%%%%%%%%%%%%%%%%%%%%%%%%%%%%
%\newpage
%\clearpage

\section{Appendix}

\begin{minipage}{\textwidth}
\begin{center}
%\begin{table*}[!t]
\begin{ruledtabular}
\begin{tabular}{>{\centering\arraybackslash}p{0.05\linewidth}>{\centering\arraybackslash}p{0.04\linewidth}>{\centering\arraybackslash}p{0.1\linewidth}>{\centering\arraybackslash}p{0.09\linewidth}>{\centering\arraybackslash}p{0.06\linewidth}>{\centering\arraybackslash}p{0.09\linewidth}>{\centering\arraybackslash}p{0.09\linewidth}>{\centering\arraybackslash}p{0.11\linewidth}>{\centering\arraybackslash}p{0.08\linewidth}>{\centering\arraybackslash}p{0.12\linewidth}>{\centering\arraybackslash}p{0.08\linewidth}}
    &$x$, \%&$|\psi_0(0)|^2$, cm${}^{-3}$& $\gamma$, rad/s/G& $F_{MW}$& $A^{theor}_s$, $\mu$eV&$A^{exp}_s$, $\mu$eV& $\langle1/r^3\rangle_p$, cm${}^{-3}$&$A^{theor}_p$, $\mu$eV&$A^{theor}_p F_{MW}$, $\mu$eV&$A^{exp}_p$, $\mu$eV\\
    \hline
    $^{207}\text{Pb}$\rule{0pt}{10pt}& 22 & $1.89\times10^{26}$& $5.58\times10^3$ & 3.07& 333.5 & 322$^a$, 126$^b$& $9.94\times10^{25}$ & 18.2 & 55.8 & 50.4$^b$\\
    \hline
    
    \multirow{2}{*}{$^{127}\text{I}$} & \multirow{2}{*}{100} & \multirow{2}{*}{$1.98\times10^{26}$}& \multirow{2}{*}{$5.39\times10^3$} & \multirow{2}{*}{1.56} & 171.3  & \multirow{2}{*}{264$^c$}& \multirow{2}{*}{$1.28\times10^{26}$} & \multirow{2}{*}{22.6} & \multirow{2}{*}{35.2} & \multirow{2}{*}{27.2$^d$}\\
    &&&&&135.5$^c$&&&&&\\
    \hline
    $^{79}\text{Br}$\rule{0pt}{10pt}& 50 & $1.65\times10^{26}$& $6.73\times10^3$ & 1.16& 132.5 & & $1.03\times10^{26}$ & 22.8 & 26.4 & 22.1$^d$\\
    \hline
    $^{81}\text{Br}$\rule{0pt}{10pt}& 50 & $1.65\times10^{26}$& $7.25\times10^3$ & 1.16& 142.7 & & $1.03\times10^{26}$ & 24.4 & 28.4 & 23.8$^d$\\
    \hline
    $^{35}\text{Cl}$\rule{0pt}{10pt}& 76 & $8.65\times10^{25}$& $2.62\times10^3$ & 1.02& 23.8 & & $5.66\times10^{25}$ & 4.8 & 4.9&\\
    \hline
    $^{37}\text{Cl}$\rule{0pt}{10pt}& 24 & $8.65\times10^{25}$& $2.18\times10^3$ & 1.02& 19.8 & & $5.66\times10^{25}$ & 4.0 & 4.1&\\
\end{tabular}
\end{ruledtabular}
\captionof{table}{\label{tab:hyperfineparam} Theoretical and experimental hyperfine constants and parameters used for their calculation. $F_{MW} = 1+3.76\times10^{-6} \:Z^3$ is the Mackey and Wood empirical factor \cite{mackey1970empirical}. The theoretical values of $\left|\psi_0(0)\right|^2$ and $\left\langle1/r^3\right\rangle_p$ (without superscripts) are given according to Morton and Preston \cite{morton1978atomic}. $A^{theor}_s=\frac{16}{3}\pi\hbar\gamma_N\mu_B\left|\psi_0(0)\right|^2 F_{MW}$, $A^{theor}_p=\frac{16}{3}\hbar\gamma_N\mu_B\left\langle\frac{1}{r^3}\right\rangle$. \cite{schawlow1949hyperfine}$^{a}$, \cite{hewes1973nuclear}$^{b}$, \cite{luc1975etude}$^{c}$, \cite{luc1973etude}$^{d}$.}
%\end{table*}
\end{center}
\end{minipage}

\vspace{1cm}

%\begin{table*}[h!]
\begin{minipage}{\textwidth}
\begin{center}
\begin{ruledtabular}
\begin{tabular}{>{\centering\arraybackslash}p{0.1\linewidth}>{\centering\arraybackslash}p{0.08\linewidth}>{\centering\arraybackslash}p{0.08\linewidth}>{\centering\arraybackslash}p{0.08\linewidth}>{\centering\arraybackslash}p{0.08\linewidth}>{\centering\arraybackslash}p{0.08\linewidth}>{\centering\arraybackslash}p{0.08\linewidth}>{\centering\arraybackslash}p{0.08\linewidth}>{\centering\arraybackslash}p{0.33\linewidth}}
 & $a_{11}/f$ & $a_{12}/f$ & $a_{44}/f$ & $C_1/f$ & $C_2/f$ & $C_3/f$ & $C_4/f$ & $f$, mT \\[1pt]
\hline
 Pb & 1 & 0 & 1 & 0 & 0 & 0 & 0 & $A_{Pb}^{p}{{\left| {{C}_{cp}} \right|}^{2}}{{x}_{Pb}}/\left( {{\mu }_{B}}{{g}_{e}} \right)$\rule{0pt}{12pt} \\[3pt]
 \hline 
I, $I_{QS}$ & 1 & 1 & 0 & 0 & 0 & 0 & 0 &  $26{{\left| {{C}_{cs}} \right|}^{2}}A_{hal}^{s}/\left( 3{{\mu }_{B}}{{g}_{e}} \right)$\rule{0pt}{12pt}\\[3pt]
\hline
I, $I_{QB}$ & -1.50 & 0.15 & -0.07 & 3.65 & -0.09 & -0.01 & -1.17 & ${{\left| {{C}_{cs}} \right|}^{2}}A_{hal}^{s}/\left( {{\mu }_{B}}{{g}_{e}} \right)$\rule{0pt}{12pt} \\[3pt]
\hline
I, $I_{1/2}$ & 1.06 & -0.03 & 0.85 & -0.15 & 0.07 & -1.94 & -0.78 & ${{\left| {{C}_{cs}} \right|}^{2}}A_{hal}^{s}/\left({{\mu }_{B}}{{g}_{e}} \right)$\rule{0pt}{12pt} \\ [3pt]
\hline
Br/Cl, $I_{QS}$ & 1 & 1 & 0 & 0 & 0 & 0 & 0 &  ${3{\left| {{C}_{cs}} \right|}^{2}}A_{hal}^{s}/\left( {{\mu }_{B}}{{g}_{e}} \right)$\rule{0pt}{12pt}\\[3pt]
\hline
Br/Cl, $I_{QB}$ & -0.28 & 0.05 & -0.25 & 1.66 & -0.03 & -0.01 & -0.82 & ${{\left| {{C}_{cs}} \right|}^{2}}A_{hal}^{s}/\left( {{\mu }_{B}}{{g}_{e}} \right)$\rule{0pt}{12pt} \\[3pt]
\hline
Br/Cl, $I_{1/2}$ & 1.53 & -0.01 & 0.72 & -0.09 & 0.04 & -1.75 & -0.94 & ${{\left| {{C}_{cs}} \right|}^{2}}A_{hal}^{s}/\left( {{\mu }_{B}}{{g}_{e}} \right)$\rule{0pt}{12pt} \\[3pt]
\end{tabular}
\end{ruledtabular}
\captionof{table}{\label{tab:coeffCondband}Coefficients connecting the components of the Overhauser field with the components of the external field and the carrier mean spin: conduction band.}
%\end{table*}
\end{center}
\end{minipage}

\vspace{1cm}
%
%
%\begin{table*}[h!]
\begin{minipage}{\textwidth}
\begin{center}
\begin{ruledtabular}
\begin{tabular}{>{\centering\arraybackslash}p{0.1\linewidth}>{\centering\arraybackslash}p{0.08\linewidth}>{\centering\arraybackslash}p{0.08\linewidth}>{\centering\arraybackslash}p{0.08\linewidth}>{\centering\arraybackslash}p{0.08\linewidth}>{\centering\arraybackslash}p{0.08\linewidth}>{\centering\arraybackslash}p{0.08\linewidth}>{\centering\arraybackslash}p{0.08\linewidth}>{\centering\arraybackslash}p{0.33\linewidth}}
 & $a_{11}/f$ & $a_{12}/f$ & $a_{44}/f$ & $C_1/f$ & $C_2/f$ & $C_3/f$ & $C_4/f$ & $f$, mT \\[1pt]
\hline
 Pb & 1 & 0 & 1 & 0 & 0 & 0 & 0 & $A_{Pb}^{s}{{\left| {{C}_{vs}} \right|}^{2}}{{x}_{Pb}}/\left( {{\mu }_{B}}{{g}_{h}} \right)$\rule{0pt}{12pt} \\[3pt]
 \hline
I, $I_{QS}$ & 1 & 1 & 0 & 0 & 0 & 0 & 0 &  $13{{\left| {{C}_{vp}} \right|}^{2}}A_{hal}^{p}/\left( 5{{\mu }_{B}}{{g}_{h}} \right)$\rule{0pt}{12pt}\\[3pt]
\hline
I, $I_{QB}$ & -3.23 & 0.32 & -0.15 & 7.4 & -0.18 & -0.05 & -4.36 & $3{{\left| {{C}_{vp}} \right|}^{2}}A_{hal}^{p}/\left( 20{{\mu }_{B}}{{g}_{h}} \right)$\rule{0pt}{12pt} \\[3pt]
\hline
I, $I_{1/2}$ & 1.78 & -0.05 & 1.59 & -0.15 & 0.14 & -2.35 & -1.67 & ${3{\left| {{C}_{vp}} \right|}^{2}}A_{hal}^{p}/\left( 20{{\mu }_{B}}{{g}_{h}} \right)$\rule{0pt}{12pt} \\[3pt]
\hline
Br/Cl, $I_{QS}$ & 1 & 1 & 0 & 0 & 0 & 0 & 0 &  $9{{\left| {{C}_{vp}} \right|}^{2}}A_{hal}^{p}/\left( 10{{\mu }_{B}}{{g}_{h}} \right)$\rule{0pt}{12pt}\\[3pt]
\hline
Br/Cl, $I_{QB}$ & -1.09 & 0.13 & -0.56 & 3.73 & -0.08 & -0.03 & -3.11 & $3{{\left| {{C}_{vp}} \right|}^{2}}A_{hal}^{p}/\left( 20{{\mu }_{B}}{{g}_{h}} \right)$\rule{0pt}{12pt} \\[3pt]
\hline
Br/Cl, $I_{1/2}$ & 2.15 & -0.02 & 1.45 & 0.15 & 0.09 & -2.18 & -2.18 & $3{{\left| {{C}_{vp}} \right|}^{2}}A_{hal}^{p}/\left( 20{{\mu }_{B}}{{g}_{h}} \right)$\rule{0pt}{12pt} \\[3pt]
\end{tabular}
\end{ruledtabular}
\captionof{table}{\label{tab:coeffValenceband}Coefficients connecting the components of the Overhauser field with the components of the external field and the carrier mean spin: valence band.}
\end{center}
\end{minipage}

\suppressfloats

\end{document}